\documentclass[aps,prb,twocolumn,nofootinbib,superscriptaddress]{revtex4-1}
\usepackage{amsfonts}
\usepackage{amscd}
\usepackage{amsmath}
\usepackage{amssymb}
\usepackage{txfonts}
\usepackage{here}
\usepackage[dvipdfmx]{graphicx}
\usepackage[dvipdfmx]{color}
\usepackage{dcolumn}
\usepackage{bm}
\usepackage{xcolor}
\usepackage[%
  colorlinks=true,
  urlcolor=blue,
  linkcolor=blue,
  citecolor=blue
]{hyperref}
\usepackage{tabularx}
\usepackage{etoolbox}
\usepackage{braket}
\usepackage[caption=false]{subfig}
\newcommand{\nn}{\nonumber}
\newcommand{\beq}{\begin{equation}}
\newcommand{\eeq}{\end{equation}}

\makeatletter
\let\cat@comma@active\@empty
\makeatother
\begin{document}

\title{
{\bf \textit{Ab initio}}
 Simulation of Non-Abelian Braiding Statistics in Topological Superconductors
}

\author{Takumi Sanno}
\email{sanno@blade.mp.es.osaka-u.ac.jp}
\affiliation{Department of Materials Engineering Science, Osaka University, Toyonaka, Osaka 560-8531, Japan}
\author{Shunsuke Miyazaki}
\affiliation{Department of Materials Engineering Science, Osaka University, Toyonaka, Osaka 560-8531, Japan}
\author{Takeshi Mizushima}
\email{mizushima@mp.es.osaka-u.ac.jp}
\affiliation{Department of Materials Engineering Science, Osaka University, Toyonaka, Osaka 560-8531, Japan}
\author{Satoshi Fujimoto}
\email{fuji@mp.es.osaka-u.ac.jp}
\affiliation{Department of Materials Engineering Science, Osaka University, Toyonaka, Osaka 560-8531, Japan}
\affiliation{Center for Quantum Information and Quantum Biology, Institute for Open and Transdisciplinary Research Initiatives, Osaka University, Toyonaka, Osaka 560-8531, Japan}
\date{\today}

\begin{abstract}
We numerically investigate non-Abelian braiding dynamics of vortices in two-dimensional topological superconductors, such as $s$-wave superconductors with Rashba spin-orbit coupling. Majorana zero modes (MZMs) hosted by the vortices constitute a topological qubit, which offers a fundamental building block of topological quantum computation. As the MZMs are protected by $\mathbb{Z}_2$ invariant, however, the Majorana qubit and quantum gate operations may be sensitive to intrinsic decoherence caused by MZM hybridization. Numerically simulating the time-dependent Bogoliubov-de Gennes equation without assuming {\it a priori} existence of MZMs, we examine quantum noises on the unitary operators of non-Abelian braiding dynamics due to interactions with neighboring MZMs and other quasiparticle states. We demonstrate that after the interchange of two vortices, the lowest vortex-bound states accumulate the geometric phase $\pi/2$, and errors stemming from dynamical phases are negligibly small, irrespective of interactions of MZMs. Furthermore, we numerically simulate the braiding dynamics of four vortices in two-dimensional topological superconductors, and discuss an optimal braiding condition for realizing the high performance of non-Abelian statistics and quantum gates operations of Majorana-based qubits.
%, and the quantum NOT gate is efficiently implemented. We discuss the conditions of braiding vortices for realizing the fundamental operations of topological quantum computation.
\end{abstract}

\maketitle
\section{Introduction}
A paramount challenge for realizing quantum computers is physical implementations of fault-tolerant quantum computation because a coupling of a quantum state to environment gives rise to unavoidable decoherence. It has been proposed that a topological phase of matter hosting non-Abelian anyons provides the hardware constitution of fault-tolerant quantum computation~\cite{kitaev06,nayakRMP08}, where topologically protected anyons lead to degenerate ground states unaffected by local perturbations and braiding such anyons implement noise-free quantum gate operations.

A Majorana fermion is a self-Hermitian relativistic particle which is equivalent to its own antiparticle~\cite{majorana}. Such fermion emerges as a special kind of Bogoliubov quasiparticles bound at defects in topological superconductors, such as vortices and edges. The observations of Majorana zero modes (MZMs) have been reported in superfluid $^3$He~\cite{mur09,murakawaJPSJ11,mizushimaJPSJ16}, unconventional superconductors~\cite{jiao20,Ran684}, superconducting nanowires~\cite{mourik12,das12,rok12,finck13,chu13,deng16,alb16,suominen17,nichele17,chen17,she17,zhang18,deng18,zanten20}, ferromagnetic atomic chains~\cite{nadj14}, quantum anomalous Hall insulator-superconductor junction~\cite{he17,Kayyalha20}, planar Josephson junctions~\cite{fornieri}, and so on~\cite{Wang52,Menard2019}. Recently, Machida {\it et al.} developed a dilution-refrigerator based STM working below 90 mK, which uncovered the existence of the zero energy vortex bound states in the iron-base superconductor Fe(Se,Te)~\cite{Machida2019}. Similar signals of Majorana bound states have also been observed in Refs.~\cite{Wang333}.
The MZMs are protected by a topological invariant. The existence of $2n$ MZMs leads to topologically protected ground states which span the $2^{n-1}$ dimensional Hilbert space and can be utilized as topological qubits. When MZMs are well-isolated from other Bogoliubov quasiparticles, they behave as non-Abelian anyons obeying the non-Abelian statistics, {\it i.e.}, braiding MZMs can implement quantum gates to manipulate the topological qubit. This is a unique character of MZMs, and topological superconductors with MZMs can provide a platform for realizing topological quantum computation~\cite{nayakRMP08,Leijnse_2012,Alicea_2012,beenakker,Stanescu_2013,elliott,sarma,satoJPSJ16,pachos,aguado2017}.

\begin{figure}[t!]
    \includegraphics[width=70mm]{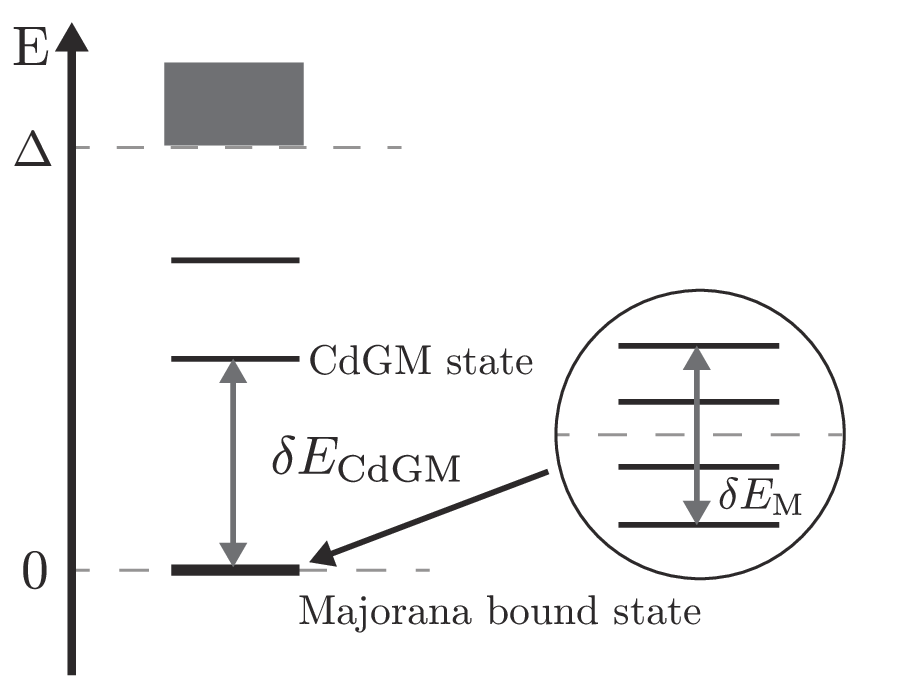}
    \caption{Schematic of splitting MZMs embedded in quasiparticle excitation spectrum. The splitting of MZMs and the level spacing from the higher CdGM states are denoted by $\delta E_{\rm M}$ and $\delta E_{\rm CdGM}$, respectively. These determine the lower and upper bounds of the time-scale of braiding dynamics: adiabatic condition, $T^{-1} << \delta E_{\rm CdGM}$, and nonadiabatic transitions between the energy levels of the hybridized MZMs}, $T^{-1} >> \delta E_{\rm M}$.
    \label{fig:adiabatic}
\end{figure}

MZMs in class D topological superconductors are, however, protected by $\mathbb{Z}_2$ topological invariant~\cite{teoPRB82,shiozakiPRB14}. %This means that in the case of class D topological superconductors, a single qubit composed of four MZMs may be fragile. 
When two neighboring vortices approach, MZM hybridization between neighboring MZMs lifts the degeneracy from zero energy to $E_{+}$  and $E_{-}$, and a pair of Majorana bound states smoothly connects to topologically trivial vortex-bound states. Thus, a single qubit composed of four MZMs may be sensitive to intrinsic decoherence caused by quasiparticle hybridization~\cite{chengPRL103,mizushimaPRA82}. The period of braiding operation, $T$, may satisfy the characteristic time scale~\cite{nayakRMP08}
\beq
\delta E _{\rm CdGM} ^{-1} \ll T \ll \delta E^{-1}_{\rm M}.
\label{eq:time}
\eeq
In this paper, we set $\hbar = 1$.
The lower bound,  $\delta E _{\rm CdGM} ^{-1}$, is to avoid the non-adiabatic transition of MZMs to higher-energy vortex-bound states~\cite{matsumoto,takigawa01,TewariPRL98,mizushimaPRL08,kraus08,kraus09,mizushimaPRA10}, {\it i.e.}, the Caroli-de Gennes-Matricon (CdGM) states~\cite{CdGM}, where the typical level spacing of vortex bound states is $\delta E _{\rm CdGM} \sim \Delta^2_0/E_{\rm F}$, where $\Delta _0$ and $E_{\rm F}$ are the bulk superconducting gap and the Fermi energy, respectively (see Fig.~\ref{fig:adiabatic}). 
The upper bound of the braiding period is associated with the hybridization of neighboring MZMs, which leads to the splitting and oscillation of the ground state energies as a function of the inter-Majorana distance $R$ as $\delta E_{\rm M} \equiv E_+-E_- \propto \cos(k_{\rm F}R)e^{-R/\xi}$, where the Fermi wavelength $k_{\rm F}^{-1}$ and the superconducting coherence length $\xi = k_{\rm F}/m\Delta _0$ determine the scale of the oscillation and splitting.  
%When the system has chiral symmetry \( \mu = 0 \), Majorana zero modes are characterized by $\mathbb{Z}$ topological invariant, which leads to that splitting energy must be zero~\cite{,chengPRB82}:  $\delta E=0$. 
 %The upper bound of the braiding period $T$ becomes infinite, but it's difficult to set experimental conditions like chiral symmetry \( \mu = 0 \). Thus, the upper bound of the braiding period $T$ is finite. $\delta E _{\rm CdGM} ^{-1}$ is the lower bound of the braiding period $T$ and the non-adiabatic effect. 
%In real systems, Majorana zero modes in vortex core are characterized by $\mathbb{Z}_2$ topological invariant, then the robustness of Majorana qubit is {\it not} trivial.
Although numerical simulation of non-Abelian statistics has been demonstrated in one-dimensional superconducting nanowires~\cite{amorimPRB91,sekania17,bauer18,andrzejPRB20,xie20,KarzigPRB91,KarzigPRX3,KnappPRX6,TutschkuPRB102}, the {\it ab initio} simulation of non-Abelian braiding dynamics and phase accumulation in two-dimensional class D superconductors has been lacking. 
In contrast to nanowire systems, where hybridization between MZMs in different wires is suppressed by gate potentials,
MZMs in vortices in two-dimensional systems are not immune to the hybridization effect. 
MZM hybridization and non-adiabatic dynamics of MZMs may disturb the non-Abelian statistics and give rise to intrinsic decoherence of Majorana qubits. %Here we discuss the performance of the unitary operators of non-Abelian braiding dynamics, based on {\it ab initio} simulations. 
It is indispensable to clarify an optimal braiding protocol for realizing the high performance of non-Abelian braiding dynamics, based on {\it ab initio} simulations. Understanding the impact of such intrinsic decoherences on Majorana-based qubits is a pressing issue. 

%------------------------------------
%-----------------------------------/

In this paper, we present {\it ab initio} simulation of non-Abelian statistics in two-dimensional topological superconductors, and numerically study intrinsic decoherence caused by quasiparticle hybridization and non-adiabatic braiding dynamics. 
%Here we examine the effects of quantum interference of neighboring MZMs and non-adiabatic braiding dynamics in class-D topological superconductors. 
By numerically simulating the time-dependent Bogoliubov-de Gennes (TDBdG) equation without assuming {\it a priori} existence of MZMs, we demonstrate the non-Abelian statistics of quantized vortices in two-dimensional trijunction network of $s$-wave superconductors with Rashba spin-orbit coupling~\cite{fujimotoPRB77,satoPRL103,satoPRB82} and the Fu-Kane model~\cite{fuPRL100}. 
We note that in addition to class D topological superconductors, our numerical method is generalizable to other topological classes which may host symmetry-protected multiple MZMs.
%Our numerical simulation is convenient. The BdG Hamiltonian used in the paper can be easily applied to various systems such as the nanowire, etc.
We first show that the particle-hole symmetry prohibits the direct transition between particle-hole symmetric MZMs. This ensures that after the interchange of two vortices, the ground state acquires the nontrivial geometric phase $\pi/2$, irrespective of interactions of MZMs. Such transition rule and geometric phase in the braiding dynamics of a two-vortex system are confirmed by numerically solving the TDBdG equation. We succeeded in extracting the geometric phase and dynamical phases from {\it ab initio} simulations and evaluating quantum noises on the braiding operators, {\it i.e.}, the unitary operators of quantum gates.
%The phase difference between initial state and final state include the nontrivial geometric phase and the dynamical phase, but the dynamical phase can't be estimated on the point of geometric symmetry of system. 
Furthermore, we perform numerical simulations of braiding dynamics in a four-vortex system, which constitutes a single topological qubit. The numerical simulations with an optimal braiding period clearly demonstrate that an initially encoded ground state is transferred to another nearly degenerate ground state by interchanging two vortices, i.e., the demonstration of non-Abelian braiding statistics with high accuracy. On the other hand, if the braiding period approaches the upper bound, the dynamical phase stemming from the splitting of ground states causes serious quantum errors of non-Abelian braiding dynamics and quantum gate operations. We discuss the upper and lower bounds of the time scale for realizing the high performance of non-Abelian braiding statistics and quantum gate operations of Majorana-based qubits. %It turns out that \red{the main factor to is energy label of Caroli-de Gennese-Matricon (CdGM) state and the distance between MZMs [???].} Here we conclude 
%We find that the period condition Eq.~\eqref{eq:time} is not the unique condition for realizing non-Abelian statistics.

The organization of this paper is as follows. In Sec.~II, we describe the theoretical framework and numerical method of the TDBdG equation and discuss the impact of the particle-hole symmetry on nonadiabatic transition between the energy levels of hybridized MZMs. %This result corresponds to Majorana effective model and topological aspect. 
We also present the  Hamiltonian relevant to  class-D topological superconductors, and present a protocol to implement the braiding dynamics of vortex singularities in superconducting trijunction systems.
In Sec.~III, we show the numerical results of braiding dynamics in two-vortex systems, and compute the geometric phase that vortex-bound states accumulate after the interchange of vortices. %We demonstrate that the time-evolution of the MZM obeys transition rule associated with the particle-hole symmetry, and they acquire the geometric phase $\pi/2$, irrespective of the quantum interference. 
In Sec.~IV, we investigate the case of four-vortex systems and present the numerical simulation of braiding dynamics in the trijunction network hosting MZMs.
Section V is devoted to conclusions and discussions on disturbance of non-Abelian statistics and decoherence of Majorana-based qubits due to  hybridization and nonadiabaticty. %A summary is given in Sec.~VI. 
The numerical results of braiding dynamics in superconductor-topological insulator heterostructures, {\it i.e.}, the Fu-Kane model, are presented in Appendix A.
%\red{The Fu-Kane model maintains the chiral symmetry as well as the particle-hole symmetry when $mu=0$. At the chiral symmetric limit, the index theorem ~\cite{WeinbergPRD24,FukuiJPSJ} ensures that a vortex with vorticitiy N hosts N zero-energy bound states, and the chiral symmetry prevents MZMs from hybridization.}
%In Appendix B, we present the summary of Majorana braiding dynamics in the effective model, where MZMs are well isolated from the other quasiparticle states, but the quantum interference is not negligible. 

\section{Time-dependent Bogoliubov-de Gennes equation}
\label{sec:2}

The dynamics of quasiparticles in superconductors is governed by the TDBdG equation. In this section we first summarize the basic properties of MZMs in class D topological superconductors. Employing the adiabatic approximation and utilizing the Majorana conditions and the particle-hole symmetry, we derive the selection rules for transition between Bogoliubov quasiparticles. We also present trijunction systems formed by $s$-wave superconductors with Rashba spin-orbit interaction. The numerical results in  superconductor-topological insulator heterostructures, {\it i.e.}, the Fu-Kane model, are shown in Appendix~\ref{sec:fu-kane}, where the impact of the chiral symmetry on the braiding dynamics is emphasized. %We also describe the 
%Previous research~\cite{chengPRB84} shows a non-negligible error has been introduced by the energy splitting and Berry connection.
%It's worth to discuss the state transition between Majorana zero modes from the point of symmetry.
%

\subsection{TDBdG equation}

We start to introduce the Bogoliubov quasiparticle operator $\eta_n(t)$ as
\beq
\hat{\eta} _n(t) = \sum _{\bm i} \left[ 
u^{\ast}_{n, {\bm i}} (t) \hat{c} _{\bm i} + v^{\ast}_{n, {\bm i}}(t)  \hat{c}^{\dag}_{\bm i} \right], 
\label{eq:etat}
\eeq
where $\hat{c}_{\bm i}$ and $\hat{c}^{\dag}_{\bm i}$ are the annihilation and creation operators of electrons at a site ${\bm i}$. The time-evolution of the quasiparticle operator is governed by the wave functions $u_{n, {\bm i}} (t)$ and $v_{n, {\bm i}} (t)$. The TDBdG equation to describe the time-evolution of quasiparticles is derived from the equation of motion for $\hat{\eta}_n(t)$. We suppose that at $t=0$, the system is in equilibrium. The corresponding quasiparticle energy $E_n$ and wave functions $\ket{\varphi_{n,{\bm i}}}\equiv[u_{n,{\bm i}},v_{n,{\bm i}}]^{\rm tr}= [u_{n,{\bm i}}(0), v_{n,{\bm i}}(0)]^{\rm tr}$ ($a^{\rm tr}$ is the transpose of a matrix $a$) are obtained from the BdG equation
\beq
\mathcal{H}_{\rm BdG}(0)  \ket{\varphi_{n,{\bm i}}}
=E_n\ket{\varphi_{n,{\bm i}}},
\eeq 
where ${\bm i}\equiv(i_x,i_y)$ and $n$ denote a site on the two-dimensional square lattice and the label of the quasiparticle energy $E_n$, respectively. 
The BdG Hamiltonian is given by
\begin{eqnarray}
\left[\mathcal{H}_{\rm BdG}(t)\right]_{{\bm i}{\bm j}} = \begin{pmatrix}
\varepsilon _{{\bm i}{\bm j}}( t) & \Delta _{{\bm i}{\bm j}}( t) \\
\Delta^{\dag}_{{\bm i}{\bm j}} (t) & - \varepsilon^{\ast}_{{\bm i}{\bm j}}( t)
\end{pmatrix},
\end{eqnarray}
where $\varepsilon (t)$ is the single-particle Hamiltonian density and $\Delta(t)$ is the pair potential. 
The Hamiltonian of class D topological superconductors obeys only the particle-hole symmetry,
\begin{equation}
	\mathcal{C} \mathcal{H}_{\rm BdG} (t)\mathcal{C}^{-1} = -\mathcal{H}_{\rm BdG} (t),
	\label{eq:phs}
\end{equation}
where \( \mathcal{C} \) is particle hole operator with $\mathcal{C}^2=+1$. This implies that the quasiparticle state with a positive energy $E_n>0$ is accompanied by the negative energy state with $-E_n$, and the quasiparticle operator obeys  
\beq
\hat{\eta}_{E_n}^{\dag}(t) = \hat{\eta}_{-E_n}(t).
\eeq
The time-evolution of the Bogoliubov quasiparticle operators in Eq.~\eqref{eq:etat} is governed by the TDBdG equation for the quasiparticle wave functions 
\beq
\label{eq:TBdG}
i\frac{\partial}{\partial t}\ket{\psi(t)}=\mathcal{H}_{\rm BdG}(t)\ket{\psi(t)}.
\eeq
The state vector is defined in the particle-hole space as $\ket{\psi(t)}_i \equiv [u_i(t), v_i(t)]^{\rm tr}$, where we impose the initial condition, $\ket{\psi(t=0)}=\ket{\varphi _{n}}$.

\subsection{Majorana fermions in class D superconductors}

Owing to the particle-hole symmetry, the zero energy quasiparticle is the equal superposition of the particle and hole components with $u^{\ast}_{E=0}=v_{E=0}$
\begin{eqnarray}
\hat{\gamma} \equiv \hat{\eta} _{E=0} = \sum _i 
u^{\ast}_{E=0, i}  \hat{c}^{}_i  + {\rm h.c.},
\end{eqnarray}
which obeys the Majorana condition, $\hat{\gamma}=\hat{\gamma}^{\dag}$.
The zero-energy quasiparticle states appear in quantized vortices of class D topological superconductor, which are protected by $\mathbb{Z}_2$ topological number~\cite{teoPRB82,shiozakiPRB14}.
This implies that the pairwise zero modes are gapped out by the hybridization of wavefunctions. Suppose class D topological superconductors with $2N$ vortices hosting $2N$ MZMs, which are well separated from other quasiparticle states. The BdG Hamiltonian then reduces to the tight-binding model composed of $2N$ MZMs,
\beq
 \mathcal{H}_{\rm eff} = i\sum_{\langle i,j\rangle} J_{ij}  \hat{\gamma}_{i} \hat{\gamma}_{j},
 \label{eq:HMF}
\eeq
where the Majorana operators bound at $i$th vortex, $\hat{\gamma}_i$, obey $\{\hat{\gamma}_i,\hat{\gamma}_j\}=\delta _{ij}$.
The hopping energy $J_{ij}$ corresponds to the energy splitting of MZMs due to hybridization~\cite{chengPRL103,mizushimaPRA82} 
\beq
  J_{ij} \propto \frac{\cos{ \left(k_{\rm F} R_{ij} + \alpha \right)}}{\sqrt{k_{\rm F} R_{ij}}} \exp{ \left(-\frac{R_{ij}}{\xi} \right)},
\label{eq:split}
\eeq
where $2\alpha=\arctan(k_{\rm F} \xi)$.
The rapid oscillation of MZMs in the scale of the Fermi wavelength $k_{\rm F}^{-1}$ represents MZM hybridization and the envelope of $J_{ij}$ is determined by the superconducting coherence length $\xi = k_{\rm F}/m\Delta_0$ and the distance between $\gamma_{i}$ and $\gamma_{j}$, $R_{ij}$. %$\xi$ is coherence length and $\alpha$ satisfy $2\alpha=\arctan(k_F \xi)$. 
This hybridization gives rise to finite splitting of ground state degeneracy. 

In this paper, however, we discuss the non-Abelian statistics of Bogoliubov quasiparticles from the direct simulation of Eq.~\eqref{eq:TBdG}. The numerical simulation of the TDBdG equation deals with the dynamics of Bogoliubov quasiparticles, {\it i.e.}, complex fermions, rather than Majorana fermions $\hat{\gamma}_j(t)$. To demonstrate the non-Abelian braiding dynamics without assuming {\it a priori} existence of MZMs, we have to obtain the Majorana fermions, $\hat{\gamma}^{1}_n(t)$ and $\hat{\gamma}^{2}_n(t)$, at the energy eigenstate $E_n$ from the Bogoliubov quasiparticles $\eta_n(t)$, and map the time-evolution of $\eta_n(t)$ onto the Majorana braiding dynamics $\hat{\gamma}^{1}_n(t)$ and $\hat{\gamma}^{2}_n(t)$. For this purpose, we decompose a spinless complex fermion $\hat{\eta}_n(t)$ associated with the energy eigenstate $E_n$ into a pair of Majorana fermions, $\hat{\gamma}^1_{n}$ and $\hat{\gamma}^2_{n}$, as 
\begin{gather}
\hat{\gamma}^{1}_{n} (t) = \frac{1}{\sqrt{2}}\left[\hat{\eta}_n(t) + \hat{\eta}^{\dag}_n(t)\right],
\label{eq:MF1} \\
\hat{\gamma}^2_{n} (t) = -\frac{i}{\sqrt{2}}\left[\hat{\eta}_n(t) - \hat{\eta}^{\dag}_n(t)\right].
\label{eq:MF2}
\end{gather}
The Majorana operators obey 
\beq
\{\hat{\gamma}^{i}_n(t),\hat{\gamma}^{j}_m(t)\}=\delta_{ij}\delta_{nm},
\eeq
($i,j = 1,2$). In general, a system with $N$ complex fermions can be represented by $2N$ Majorana fermions $\{ \hat{\gamma}^{i}_n \}^{i=1,2}_{n=1,\cdots,N}$. The $N$ complex fermions allow one to introduce $2^N$ dimensional Fock states, 
\beq
\ket{a_1a_2\cdots a_N}=\bigotimes_{n=1}^N \ket{a_n}
\equiv \bigotimes_{n=1}^N (\hat{\eta}^{\dag}_n)^{a_n}\ket{0},
\label{eq:fock}
\eeq
where $a_n=\{0, 1\}$ is the occupation number of the quasiparticle state with $E_n$. The eigenvector $\ket{a_n=0}$ is the vacuum of the Bogoliubov quasiparticle with an eigenenergy $E_n$ and $\ket{1}=\hat{\eta}_n^{\dag}\ket{0}$ is the occupied state. The $2^N$ degenerate ground states are lifted by the tunneling splitting of $2N$ MZMs. We also note that although the BdG Hamiltonian violates the particle number conservation, the particle-hole symmetry ensures the conservation of the fermion parity. Hence, the Fock state is an eigenvalue of the parity operator $\hat{P} = (-1)^{\hat{F}}$, where $\hat{F}=\sum _{{\bm i},\sigma}c^{\dag}_{{\bm i},\sigma}c_{{\bm i},\sigma}$ is the fermion number operator. The fermion parity conservation splits the Fock space to the $2^{N-1}$-dimensional subsectors of the fermion parity. 
In the limit of $E_n\rightarrow 0$, the Majorana operator $\hat{\gamma}^{i}_n$ reduces to the MZM $\hat{\gamma}_i$ bound at a vortex core.

Let us consider a system with $2N$ vortices. When the vortices are well isolated from each other, they host $N$ particle-hole symmetric pairs of zero-energy Bogoliubov quasiparticles. The inter-vortex tunneling of the quasiparticles leads to the dispersive band-like structure with the band width $\delta E_{\rm M}$. The interchange of vortex-bound Bogoliubov quasiparticles can be implemented by braiding vortices with a time period $T$. 
As shown in Fig.~\ref{fig:adiabatic}, vortex systems in class D topological superconductors have two typical energy scales, the level spacing between MZMs and non-Majorana vortex states, $\delta E_{\rm CdGM} \sim \mathcal{O}(\Delta^2_0/\varepsilon_{\rm F})$, and the splitting of MZMs, $\delta E_{\rm M} \sim \cos(k_{\rm F}R)e^{-R/\xi}$. We assume that the ``Majorana band'' is well separated from higher CdGM states, i.e., $\delta E_{\rm M}\ll \delta E_{\rm CdGM}$, and the braiding operation satisfies the adiabatic regime
\beq
T \gg \delta E_{\rm CdGM}^{-1}.
\label{eq:adiabatic}
\eeq
To capture the braiding dynamics of vortex-bound Bogoliubov quasiparticles, we introduce the Majorana representation of the Bogoliubov quasiparticles in Eqs.~\eqref{eq:MF1} and \eqref{eq:MF2}, where $N$ Bogoliubov quasiparticle states can be represented by $2N$ Majorana fermions $\hat{\gamma}^i_n(t)$. The TDBdG equation \eqref{eq:TBdG} describes the unitary evolution of the Bogoliubov quasiparticles as $\hat{\eta}_n(t)\equiv U(t)\hat{\eta}_n(0)U^{\dag}(t)$. 
Within the adiabatic condition in Eq.~\eqref{eq:adiabatic}, the braiding dynamics of vortex-bound Bogoliubov quasiparticles can be regarded as the unitary time evolution of the Majorana operators, 
\beq
\hat{\gamma}^i_n(t)\equiv U(t)\hat{\gamma}^i_n(0)U^{\dag}(t).
\eeq
%period $T$ the $2N$ dimensional Majorana operators are rotated by the ${\rm SO}(2N)$ matrix ${\bf V}$ as 
%The time-evolution of the Bogoliubov quasiparticles, i.e., the unitary evolution of the operator $\eta_n(t)\equiv U(t)\eta_n(0)U^{\dag}(t)$, is governed by the TDBdG equation \eqref{eq:TBdG}.  
Following Ref.~\onlinecite{chengPRB84}, we introduce the ${\rm SO}(2N)$ matrix ${\bf V}$ which rotates the $2N$ dimensional Majorana operators as 
\beq
\hat{\gamma}^i_n(T) = \sum _{j,m}{\bf V}^{ij}_{nm}\hat{\gamma}^j_m(0),
\label{eq:braiding}
\eeq
where ${\bf V}$ is subject to the conservation of the fermion parity. An explicit expression of ${\bf V}$ is obtained from the numerical simulation of the TDBdG equation \eqref{eq:TBdG}. The braiding matrix $U(T)$ is obtained from the computed matrix ${\bf V}$ as 
\beq
U(T) = \exp \left(\frac{1}{4}\sum _{ij}{\bf D}^{ij}_{nm}\hat{\gamma}^i_n\hat{\gamma}^j_m\right),
\label{eq:braidingU}
\eeq
where $e^{-{\bf D}}={\bf V}$~\cite{chengPRB84}. In Secs.~\ref{sec:two} and \ref{sec:four}, we evaluate quantum noise effects on non-Abelian statistics by computing the ${\rm SO}(2N)$ matrix ${\bf V}$ from the TDBdG equation in class-D topological superconductors with two and four vortices, respectively.

\subsection{Nonadiabatic transition rules and particle-hole symmetry}
\label{sec:selection}

We start to derive a generic result that the PHS imposes selection rules on the transition between instantaneous eigenstates of $\mathcal{H}_{\rm BdG}(t)$. We consider a superconducting state with $2N$ vortices, where each vortex hosts a single MZM. We introduce a set of time-dependent parameters $\bm{R} (t) =(R_1(t),...,R_N(t))$ as a vector in the parameter space. Let ${\bm R}_i(t)$ be the position of the $i$th vortex. The interchange of the $i$th vortex and the $j$th vortex is implemented by exchanging ${\bm R}_i(t)$ and ${\bm R}_j(t)$ to ${\bm R}_i(T)={\bm R}_j(0)$ and ${\bm R}_j(T)={\bm R}_i(0)$, and the twice operation, ${\bm R}_i(2T)={\bm R}_i(0)$, defines a cyclic trajectory in the parameter space. 
The time dependence of the BdG Hamiltonian is described through the parameters $\bm{R} (t) $ as $\mathcal{H}_{\rm BdG}(t)\rightarrow\mathcal{H}_{\rm BdG}(\bm{R})$. Let $\ket{\varphi_n({\bm R})}$ be an instantaneous eigenstate of $\mathcal{H}_{\rm BdG}(\bm{R})$: 
\beq
\mathcal{H}_{\rm BdG}(\bm{R})\ket{\varphi_n (\bm{R})}= E_n (\bm{R})\ket{\varphi_n (\bm{R})}, 
\label{eq:BdGinst}
\eeq
where \( \ket{\varphi_n (\bm{R})} \) satisfies the orthonormal condition 
\beq
\braket{\varphi_n (\bm{R})|\varphi_m (\bm{R})}=\delta_{n,m}.
\label{eq:norm}
\eeq
We now expand the time-evolution of the $n$th eigenstate in terms of the set of instantaneous eigenstates $\{ \ket{\varphi_m (\bm{R})}\}$ as
\beq
\label{expand_state}
\ket{\psi_n(t)}=\sum_{m} C^{(n)}_m (t) e^{-i\int_0^t {E}_m (t^{\prime}) dt^{\prime}} \ket{\varphi_m (\bm{R})},
\eeq
where \( e^{-i\int_0^t E_n (t') dt'} \) is the dynamical phase and we have introduced $n\in\mathbb{Z}$ as labels of eigenstates. The state vector is assumed to obey the initial condition, $\ket{\psi_n(t=0)}=\ket{\varphi_n({\bm R}(0))}$, corresponding to 
\beq
C^{(n)}_m(t=0) = \delta_{nm}.
\eeq
Substituting Eq.~\eqref{expand_state} to Eq.~\eqref{eq:TBdG}, the equation for the coefficient \( C^{(n)}_m (t) \in \mathbb{C}\) is given as 
\begin{equation}
\label{state_eq}
i{\partial}_t C^{(n)}_m (t) + \sum_{k} \Pi_{mk} (t) e^{-i\int^t_0 \Delta E_{km} (t') dt'} C^{(n)}_k (t) = 0.
\end{equation}
The hermitian matrix, $\Pi_{mk} (t) = \Pi^{\ast}_{km} (t) $, represents the transition probability between $m$-th and $k$-th instantaneous eigenstates, which is given as
\beq
\Pi_{nm} (t) \equiv i\braket{\varphi_n (\bm{R})|{\partial}_t \varphi_m (\bm{R})} .
\eeq
Under the adiabatic approximation, $\Pi_{nm}$ reduces to an element of the Berry connection matrix.

We consider the braiding dynamics that the operation period $T$ satisfies the adiabatic approximation in Eq.~\eqref{eq:adiabatic},
%\beq
%\delta E_{\rm CdGM}\gg T^{-1},
%\eeq
which implies that the higher CdGM states are outside the ground-state (i.e., Majorana fermions) subspace. Consider a class D topological superconductor with $2N$ vortices. When the inter-vortex distance is sufficiently large, tunneling probability of quasiparticles between neighboring vortices is negligible, and each vortex hosts an exactly zero energy state. In such ideal situation, the braiding rule is governed by the matrix $\bf{D}$ and the unitary matrix in Eq.~\eqref{eq:braidingU} reduces to $U(T)=\exp(\frac{\pi}{4}\hat{\gamma}_i\hat{\gamma}_j)$, implying that the MZMs obey non-Abelian statistics. In finite size systems, however, quasiparticle tunneling between vortices always leads to a nonzero splitting of zero energy states, $\pm \delta E_{\rm M}$ as in Eq.~\eqref{eq:split}. The Bogoliubov quasiparticle operator after the interchange operation is given by substituting Eq.~\eqref{expand_state} into Eq.~\eqref{eq:etat} as 
\beq
\hat{\eta}_n(T) = \sum _{m,k}C^{(n)}_m(T)e^{-i\int^T_0 E_{m} (t) dt}B_{mk}\hat{\eta}_k(0).
\label{eq:etatn}
\eeq
Here we have introduced the matrix element $B_{mk}\equiv \braket{\varphi_m({\bm R}(0))|\varphi_k({\bm R}(T))}$, which describes the transformation of the quasiparticle basis $\hat{\eta}_n(0)$ to $\sum _m B_{nm}\hat{\eta}_m(0)$ after braiding operation. The expression of matrix $U(T)$ in Eq.~\eqref{eq:braidingU}, can be directly read off from Eq.~\eqref{eq:etatn}. Cheng {\it et al.}~\cite{chengPRB84} found that tunneling splitting of zero-energy states gives rise to the nonuniversal contributions of dynamical phase and non-Abelian Berry phase, and the resulting braiding matrix contains a non-negligible error as 
\beq
U(T) = \exp\left[\left(
\frac{\pi}{4} - \frac{\mathcal{E} T}{2}
\right)\hat{\gamma}^1\hat{\gamma}^2 \right],
\label{eq:umatrix}
\eeq
where $\mathcal{E}\sim |\delta E_{\rm M}|$. As $\mathcal{E}T/2=\mathcal{O}(1)$ for $T \gtrsim \delta E^{-1}_{\rm M}$, the noise error induced by tunneling splitting becomes significant when the braiding operation is slow compared to the time scale due to tunneling splitting of MZMs.
Such serious error can lead to bit flip error and parity error~\cite{ScheurerPRB88, chengPRB84}.

%\red{Comments on Cheng. The braiding operator $U$ }

Therefore we consider non-adiabatic transitions between splitting MZMs, where the period $T$ of braiding vortices is much faster than the time scale associated with the splitting of degenerate ground states $\delta E_{\rm M}$,  i.e., 
\begin{eqnarray}
T \ll \delta E^{-1}_{\rm M}
\label{eq:period_set}
\end{eqnarray}
As $R\gg \xi $ and $\Delta_0\gg \varepsilon_{\rm F}$ in a realistic situation, the energy scales satisfy $\delta E_{\rm CdGM} \gg \delta E_{\rm M}$, and the operation period obeys the conditions in Eqs.~\eqref{eq:adiabatic} and \eqref{eq:period_set}. %Hence, we assume the period of braiding vortices, $T$, satisfies the following conditions
%\begin{eqnarray}
%\delta E_{\rm M}\ll T^{-1}
%\label{eq:period_set}
%\end{eqnarray}
%The first condition imposes the adiabatic approximation, where MZMs are well separated from other CdGM states. The second condition admits  non-adiabatic transitions between the splitting MZMs. 
We also note that within the condition in Eq.~\eqref{eq:period_set}, $\Delta E_{nm} T \ll 1 $ and the dynamical phase due to splitting of MZMs is negligible, \(  e^{-i\int^t \Delta E_{nm} (t') dt'} \approx 1 \). Thus Eq.~\eqref{state_eq} is recast into
\begin{equation}
i \partial _t {C}^{(n)}_m (t) + \sum _k \Pi_{mk} (t)  {C}^{(n)}_k (t) = 0.
\label{eq:differ}
\end{equation}
%where we set $\bm{C} (t) \equiv [C_{+1} (t), C_{-1} (t), ... , C_{+n} (t), C_{-n} (t) )]^T$, and we have introduced the abbreviation, $\Pi (t)  \bm{C} (t)\equiv\sum_{n} \Pi_{mn} (t) C_n (t)$.
%and \( \Pi (t) \) is hermitian : \( \Pi (t) = \Pi^{\dagger} (t) \). 
%
As the differential equation for \( {C}^{(n)}_m (t) \) is determined by the Berry connection matrix \( \Pi (t) \), we clarify the roles of the symmetry and Majorana condition on the matrix $\Pi(t)$.
Let us consider Bogoliubov quasiparticles with the energy $E_n>0$ and the eigenvector $\ket{\varphi_n (\bm{R}(t))}$.
%\begin{equation}
%\ket{\varphi_n (\bm{R}(t))} = \left(\begin{array}{c} u_{E_n} (\bm{R}(t)) \\ v_{E_n} (\bm{R}(t)) \end{array}\right).
%\end{equation}
The particle-hole symmetry in Eq.~\eqref{eq:phs} ensures that the positive energy eigenstate is always accompanied by a negative energy eigenstate with $-E_n$ and $\ket{\varphi_{-E_n}}=\mathcal{C} \ket{\varphi_{+E_n}}$.
%
%the existence of the eigenstate with $-E_n$ and $\ket{\varphi_{-E_n}}=\mathcal{C} \ket{\varphi_{+E_n}}$. This indicates that a positive %energy eigenstate is always accompanied by a negative energy eigenstate, and the wave function obeys 
%\begin{equation}
%\label{PHS}
%$(u_{-E_n}, v_{-E_n})=(v^*_{+E_n}, u^*_{+E_n})$.
%\end{equation}
%The exact zero energy satisfies the Majorana condition, 
%\beq
%v_{\pm E_n}=u^*_{\pm E_n}.
%\label{eq:majorana}
%\eeq
The Berry connection matrix \( \Pi (t) \) is subject to the particle-hole symmetry in Eq.~\eqref{eq:phs} and the orthonormal condition in Eq.~\eqref{eq:norm}. %,  and the Majorana condition in Eq.~\eqref{eq:majorana}. Using the Majorana condition, one first finds that $\braket{\varphi_{\pm{E_n}} (\bm{R})|{\partial}_t \varphi_{\pm{E_n}} (\bm{R})} =  \partial_t (u_{\pm{E_n}} u_{\pm{E_n}} ^*) = 0$ and the diagonal terms vanishes, 
%\beq
%\Pi_{\pm E_n,\pm E_n} (t) = 0.
%\eeq
%The particle-hole symmetry in Eq.~\eqref{eq:phs} and the orthonormal condition in Eq.~\eqref{eq:norm} 
These lead to $\braket{\varphi_{-E_n} (\bm{R})|{\partial}_t \varphi_{+E_n} (\bm{R})}= \braket{\varphi _{-E_n}(\bm{R})|{\partial}_t \mathcal{C}\varphi_{-E_n} (\bm{R})}=\frac{1}{2}\partial _t\braket{\varphi _{-E_n}(\bm{R})|\mathcal{C}\varphi_{-E_n} (\bm{R})} = \partial _t\braket{\varphi _{-E_n}(\bm{R})|\varphi_{+E_n} (\bm{R})} =0 $. Hence, the symmetry prohibits direct transition between the particle-hole symmetric eigenstates,  
\beq
\Pi_{+ E_n,- E_n} (t) =\Pi_{- E_n,+ E_n} (t) = 0.
\label{eq:pimatrix}
\eeq
In the same manner, one reads from Eq.~\eqref{eq:phs} and \eqref{eq:norm}  
\begin{align}
\Pi_{\mp E_m, -E_n}(t)=& - \Pi_{\pm E_m,+E_n}^*(t).
\end{align}
%
%Thus, Berry matrix connection is as follow :
%\begin{align}
%\label{eq:matrix}
%\Pi =
%\begin{pmatrix} 
%\hat{0}_{2\times 2} & \hat{A}_{E_1,E_2}  & \cdots & \cdots & \hat{A}_{E_1,E_N} \\ 
%\hat{A}^{\dag}_{E_1,E_2}  & \hat{0}_{2\times 2}  & \hat{A}_{E_2,E_3}  & \cdots & \hat{A}_{E_2,E_N} \\
%\vdots &\ddots & \ddots &  \ddots & \vdots \\
%\hat{A}^\dag _{E_1,E_{N-1}} & \cdots & \hat{A}^\dag _{E_{N-2},E_{N-1}} & \hat{0}_{2\times 2}   & \hat{A} _{E_{N-1},E_N}\\
%\hat{A}^\dag _{E_1,E_N} & \cdots & \cdots & \hat{A}^\dag _{E_{N-1},E_N}   & \hat{0}_{2\times 2} 
%\end{pmatrix} ,
%\end{align}
%where \( \hat{A}_{E_n,E_m} (t) \) is
%\begin{align}
%\hat{A} _ {E_n, E_m} (t) \equiv &
%\begin{pmatrix} \Pi_{+E_n,+E_m} (t) & \Pi_{+E_n,-E_m} (t) \\ 
% -\Pi_{-E_n,+E_m} (t) & -\Pi _{-E_n,-E_m} (t) 
%\end{pmatrix} \nn \\
% =&
%\begin{pmatrix} \Pi_{+E_n,+E_m} (t) & \Pi_{+E_n,-E_m} (t) \\ 
% -\Pi^*_{+E_n,-E_m} (t) & -\Pi^* _{+E_n,+E_m} (t) 
% \end{pmatrix}  .
%\end{align}
%Equation~\eqref{eq:matrix}
The property of the Berry connection matrix described above indicates that transition between particle-hole symmetric MZMs is forbidden by the particle-hole symmetry. Such particle-hole symmetric transition violates the fermion parity. In Sec.~\ref{sec:two}, we demonstrate that the numerical simulation of braiding dynamics is consistent to the transition rules, and the prohibition of particle-hole symmetric transition ensures the acquisition of the geometric phase $\pi/2$, irrespective of the splitting of MZMs, $\delta E_{\rm M}\neq 0$. %[Comments on Cheng's work?]} %Indirect transition between pairs is possible with parity breaking. This phenomena does not show non-conserving fermion parity. Generally, energy eigenstates in BdG Hamiltonian does not perform one-to-one correspondence to the eigenstates expanded in Majorana effective model. Qubit is desirable whether the energy eigenstate is occupied or not, for example $\ket{00}$ corresponds to occupation of $\ket{\varphi_{-E_1}}$ and $\ket{\varphi_{-E_2}}$.\red{[???]}

\subsection{$s$-wave superconductors with Rashba spin-orbit interaction}

As a model of class D topological superconductors, we consider an $s$-wave superconductor with spin-orbit interaction (SOI) ~\cite{fujimotoPRB77,satoPRL103,satoPRB82}. The Hamiltonian is given by 
\beq
\mathcal{H} = \mathcal{H}_{\rm K} + \mathcal{H}_{\rm Z} + \mathcal{H}_{\rm SOI} + \mathcal{H}_{\rm SC}.
\label{eq:rashba}
\eeq
The model is composed of the simple building blocks, such as the hopping term ($\mathcal{H}_{\rm K}$), the magnetic Zeeman term ($\mathcal{H}_{\rm Z}$), the Rashba SOI ($\mathcal{H}_{\rm SOI}$), and the $s$-wave pairing term ($\mathcal{H}_{\rm SC}$),
%
%We suppose class D topological superconductor, topological invariant is \( Z_2 \), then use two-dimentional s-wave supercondcutor system with spin-orbit interaction(SOI) ~\cite{fujimotoPRB77,satoPRB79,satoPRB82} which is called Lashba model and described by the Hamiltonian \(  \mathcal{H}_{\rm BdG} = \mathcal{H}_{\rm kin} + \mathcal{H}_{\rm SC} + \mathcal{H}_{\rm SOI} \);
\begin{align}
%\begin{split}
\mathcal{H}_{\rm K}=& -t_0 \sum_{ \left<\bm{i},\bm{j}\right>,\sigma} \hat{c}^\dag_{\bm{i},\sigma} \hat{c}^{ }_{\bm{j},\sigma}
-\mu\sum_{ \bm{i},\sigma}\hat{c}^\dag_{\bm{i},\sigma} \hat{c}^{ }_{\bm{i},\sigma} ,
\\
%&
\mathcal{H}_{\rm Z}=&-\mu_{\rm B} H_z \sum_{\bm{i},\sigma, \sigma '} (\sigma_z)_{\sigma, \sigma '}  \hat{c}^\dag_{\bm{i},\sigma} 
\hat{c}^{ }_{\bm{i},\sigma '},
\\
\mathcal{H}_{\rm SOI} =& -\lambda \sum_{\bm{i} } \left[ \left( \hat{c}^\dag _{\bm{i} - \hat{\bm x}, \downarrow} \hat{c}^{ } _{\bm{i} , \uparrow} - \hat{c}^\dag _{\bm{i} + \hat{\bm x} , \downarrow} \hat{c}^{ } _{\bm{i}, \uparrow} \right) \right.
\nn \\
& 
\left.+ i \left( \hat{c}^\dag _{\bm{i} - \hat{\bm y}, \downarrow} \hat{c}^{ } _{\bm{i}, \uparrow} 
- \hat{c}^\dag _{\bm{i} + \hat{\bm y}, \downarrow} \hat{c}^{ } _{\bm{i}, \uparrow} \right) + {\rm h.c.} \right],
\\
\mathcal{H}_{\rm SC} =& \sum_{\bm{i}} \Delta_{\rm s} e^{i \theta_{\bm{i} } } \hat{c}^\dag _{\bm{i}, \uparrow} \hat{c}^\dag _{\bm{i}, \downarrow} + {\rm h.c.}.
%\end{split}
\end{align}
Here \( \hat{c}^\dag _{\bm{i}, \sigma} \left( \hat{c}_{\bm{i}, \sigma} \right) \) is a creation (annihilation) operator of electrons with spin \( \sigma = (\uparrow,\downarrow) \) at site \( \bm{i} = (i_x, i_y) \). 
In numerical calculations, we set the parameters as $t_0=1.0$, $\mu=-6.2$, $\mu_{\rm B} H_z =5.0$, $\lambda=2.0$, and $\Delta_s=2.5$ and scaled with $t_0$. 
This choice of parameters ensures that odd number of MZMs exist in a vortex core.
%This model don't need topological insulator (TI) like fu-kane model, which says that it can be easily set as an experimental system.
%\red{[???]Moreover, MZMs state is more stable than that of p-wave superconductor when zeeman energy satisfies the condition \( 0<\Delta - \mu_{\rm B} H_z < \Delta \).~\cite{satoPRB82}}

%-------------------------------------------------
\begin{figure}
  \centering
  %\begin{tabular}{c}
  \subfloat{\includegraphics[width=4.5cm,height=4cm]{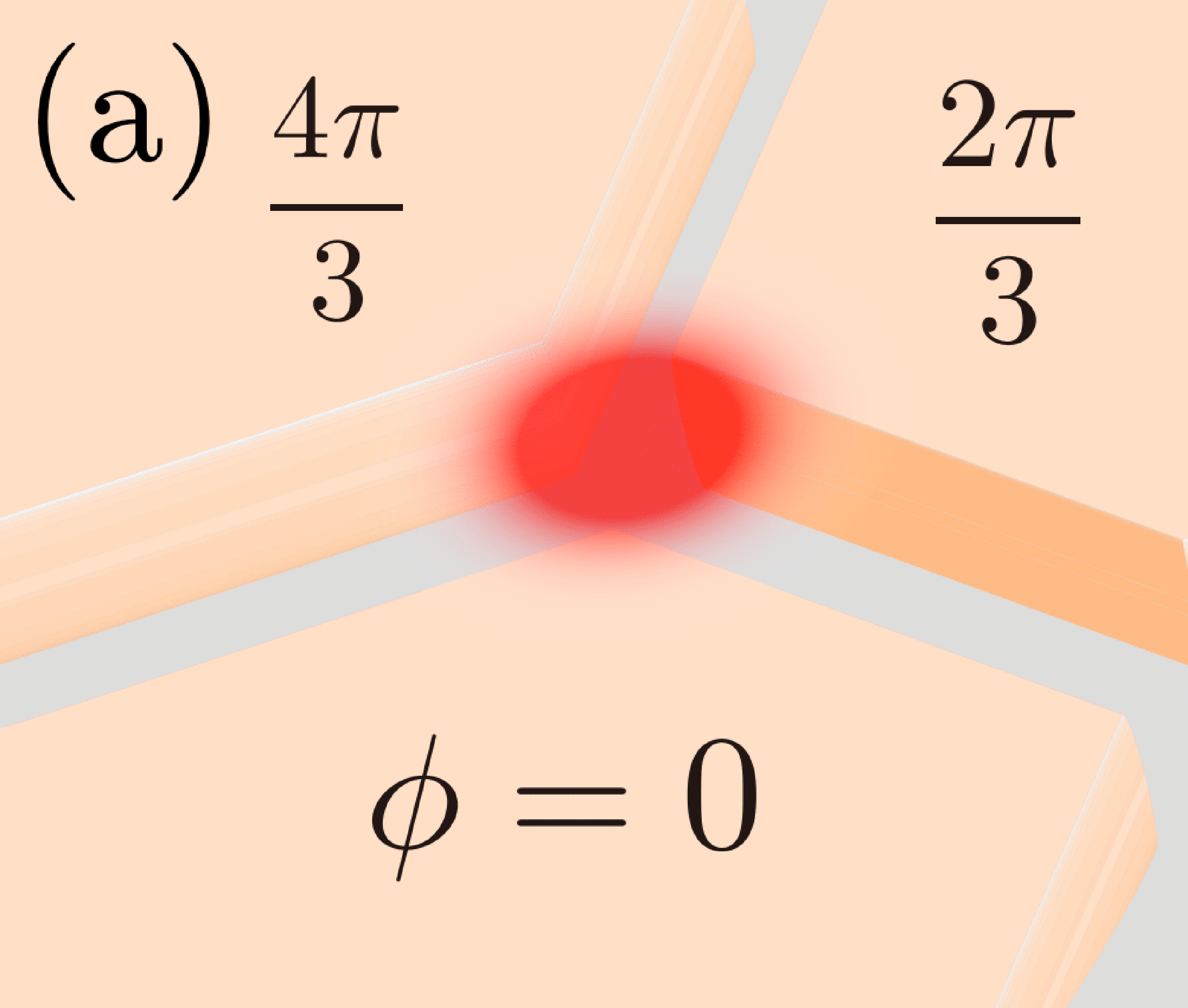}
  \label{fig:trijunction1}}
  %\hspace{0.5cm}
  \subfloat{\includegraphics[width=4cm,height=4cm]{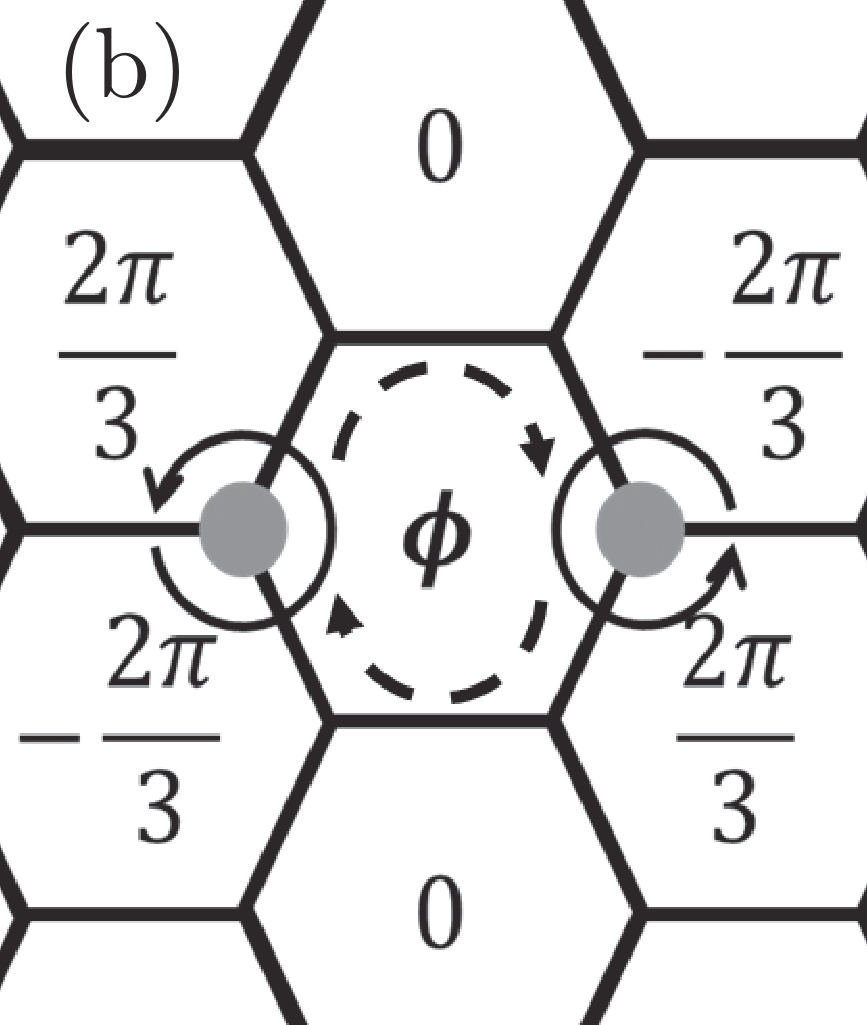}
  \label{fig:trijunction2}}
  \caption{(a) Schematic of a trijunction hosting a single MZM. Each superconducting island has different phase, $\phi = 0, \pm 2\pi/3$, and the intersection is regarded as a phase singularity, {\it i.e.}, a vortex singularity. (b) Two-dimensional network of trijunctions. The braiding of MZMs is implemented by rotating the ${\rm U}(1)$ phase \( \phi \), where the dashed lines indicate the motion of the phase singularities with changing $\phi$ from 0 to 2\( \pi \).}
   \label{trijunction}
   %\end{tauar}
\end{figure}
%------------------------------------------------/

%---
\subsection{Numerical Method}

To demonstrate the braiding dynamics of vortices and non-Abelian statistics, we numerically solve the TDBdG equation~\eqref{eq:TBdG} for two ($N=1$) and four ($N=2$) MZMs in Secs.~\ref{sec:two} and \ref{sec:four}, respectively. The time-evolution of the quasiparticle state from \( \ket{\psi(t)} \) to \( \ket{\psi(t+\Delta t )} \) is governed by the time-evolution operator \( \hat{U} (t+\Delta t;t) \),
\begin{equation}
	\ket{\psi (t+\Delta t) } = \hat{U} (t+\Delta t; t) \ket{\psi(t)}.
\end{equation}
In general, the time-evolution operator is given by
\begin{equation}
	\hat{U} (t+\Delta t;t) = \hat{T} \exp \left[-i\int_t^{t+\Delta t} dt' \mathcal{H} _{\rm BdG}(t') \right],
\end{equation}
which is approximately reduced to \( \hat{U} (t+\Delta t ;t) \approx \exp \left[ -i \mathcal{H} (t) \Delta t \right]  \) for small \( \Delta t \). 
We expand the time-evolution operator in terms of the Chebyshev polynomials~\cite{talJCP81}. The similar method is utilized for the simulation of the braiding dynamics in one-dimensional superconducting nanowires~\cite{amorimPRB91,andrzejPRB20}.
We numerically simulate the Majorana braiding dynamics in a two-dimensional network of trijunctions as shown in Fig.~\ref{trijunction}\subref{fig:trijunction1}. This was first proposed by Fu and Kane~\cite{fuPRL100}. Each superconducting island has different $U(1)$ phase, $\phi = 0, \pm 2\pi/3$, and the intersection is regarded as a phase singularity, {\it i.e.}, a vortex singularity hosting a single MZM. As shown in  Fig.~\ref{trijunction}\subref{fig:trijunction2}, the braiding operation of vortices is implemented by changing the ${\rm U}(1)$ phase \( \phi \) in a superconducting island, which is induced by ``quantum'' phase slips \( \phi \rightarrow \phi + 2\pi \). The quantum phase slip has already been observed in a superconducting nanowire~\cite{astacoherent}, and a crossover between quantum and thermal quantum slips has also been realized with changing temperatures~\cite{masudaAPL108}. The junction system can be applied as surface code for fault-tolerant quantum computation~\cite{vijayPRX5,vijay2016}.
%Now, we transform the spatial dependence of the phase as shown in Fig.\ref{phase} to simplify the numerical calculation. Additionally, we set that two vortices model Fig.\ref{phase}-a is open boundally condition and four vortices model Fig.\ref{phase}-b is periodic boundally condition imposed on the same color part.

%======================================================
\section{Braiding dynamics in two-vortex systems }
\label{sec:two}

In this section, we consider Majorana braiding dynamics in the Rashba model in Eq.~\eqref{eq:rashba} with two vortices, {\it i.e.}, $N=1$. When two vortices are well separated and host exactly zero energy states, the braiding of vortices transforms the Majorana operators $\hat{\gamma}^1$  and $\hat{\gamma}^2$ to $\hat{\gamma}^2$ and $-\hat{\gamma}^1$. The transformation, $\hat{\gamma}^j \rightarrow U(T)\hat{\gamma}^j U^{\dag}(T)$, is compactly represented by the unitary matrix in Eq.~\eqref{eq:braidingU} %, which is recast into
\beq
U(T)=\exp{\left[\vartheta(T)\hat{\gamma}^1\hat{\gamma}^2\right]},
\label{eq:umatrix2}
\eeq
with $\vartheta=\pi/4$. When tunneling splitting of MZMs is not negligible, however, the angle $\vartheta(T)$ is deviated from the ideal value by quantum noises arising from the extra contributions of the dynamical phase and Berry phase as shown in Eq.~\eqref{eq:umatrix}. Here we will clarify the conditions in which intrinsic noise effects are negligible, and $\vartheta(T)$ approaches $\pi/4$. 

Solving the TDBdG equation \eqref{eq:TBdG} without assuming {\it a priori} existence of MZMs, we obtain the information of the transformation of quasiparticles after braiding, $\hat{\eta}_n(T) = \sum _m {\bf V}_{nm}\hat{\eta}_m(0)$, where ${\bf V}_{nm}$ is composed of the transition coefficient ($C^{(n)}_m$), the dynamical phase, and the transformation of the quasiparticle basis ($B_{mk}$) as in Eq.~\eqref{eq:etatn}. We compute the transition probability of the initial state $\ket{\psi_n(0)}=\ket{\varphi_n({\bm R}(0))}$,
\beq
P^{(n)}_m(T) \equiv \left|\braket{\varphi_m ({\bm R}(0))|\psi_n(T)}\right|,
\label{eq:pnm}
\eeq
and the accumulation of the geometric phase $\phi_{\rm geo}$
\beq
\phi _{\rm geo}(T) \equiv {\rm arg}\braket{\psi_n(0)|\psi_n(T)} - \phi _{\rm dyn}(T),
\eeq
after the interchange of vortices, where $\phi _{\rm dyn}(T)$ is the dynamical phase which the wave function accumulates after the interchange of vortices. From $P^{(n)}_m(T)$ and $\phi _{\rm geo}(T)$, we obtain the explicit expression of the braiding matrix $U(T)$, and extract the effect of nonadiabaticiy and tunneling splitting on quantum noise.

%  $\vartheta(T)$ in the unitary matrix without assuming {\it a priori} existence of MZMs, we 

%When tunneling splitting of MZMs is not negligible, however, we will show below that as shown in Eq.~\eqref{eq:umatrix}, the angle $\vartheta$ is the nonadiabaticity of braiding operation within the condition in Eq.~\eqref{eq:period_set} is crucial for realizing the 

%contrary to Eq.~\eqref{eq:umatrix}. In braiding two vortices, the ground state acquires a phase shift, which is very important to confirm the consistency between the Majorana effective model and the full numerical calculation of the TDBdG equation.
%According to the Majorana effective model, 

For a class D superconductor with $2N$ vortices, as shown in Eq.~\eqref{eq:fock}, $N$ particle-hole symmetric vortex-bound states $\eta_n$ construct the $2^N$-dimensional Fock space spanned by $\ket{a_1\cdots a_N}$ ($a_j=\{0,1\}$). The $2^N$ degeneracy of the ground states are lifted by the tunneling splitting of $2N$ MZMs. 
For $N=1$, the hybridization of two MZMs leads to the energy splitting $\pm E_{+}$, where $E_{+} = J_{12}$ in Eq.~\eqref{eq:split}. Two Fock states are introduced as 
%Energy eigenstates of ground states correspond to ground states in Fock space spanned by complex fermion composed of two MZMs; 
%\begin{equation*}
the even parity state $\ket{0}$ and the odd parity state $\ket{1}\equiv \hat{\eta}_+^{\dag}\ket{0}$ with $\hat{\eta}_+^{\dag}$ being the creation operator of the energy eigenstate with $+ E_+$. These two states belong to the different parity eigenstate and thus the transition is prohibited unless the particle-hole symmetry is broken.

%Above result says fermion parity conservation for two vortices case from PHS. 
%Therefore, $\ket{1}(\ket{0})$ can't transit $\ket{0}(\ket{1})$.
%Also, these results correspond to topological results (fermion parity conservation) and keeping bonding orbit (anti-bonding orbit).

%======================================================
%two vortices setup figures
\begin{figure}
%  \centering
  %\begin{tabular}{ccc}
%  \subfloat{\includegraphics[width=7cm,height=5cm]{numerical_setup_2.png}\label{fig:setup_2}}
%  \hspace{0.5cm}
%  \subfloat{\includegraphics[width=7.5cm,height=4cm]{vortex-two_state.png}\label{fig:state_2}}
  \subfloat{\includegraphics[width=80mm]{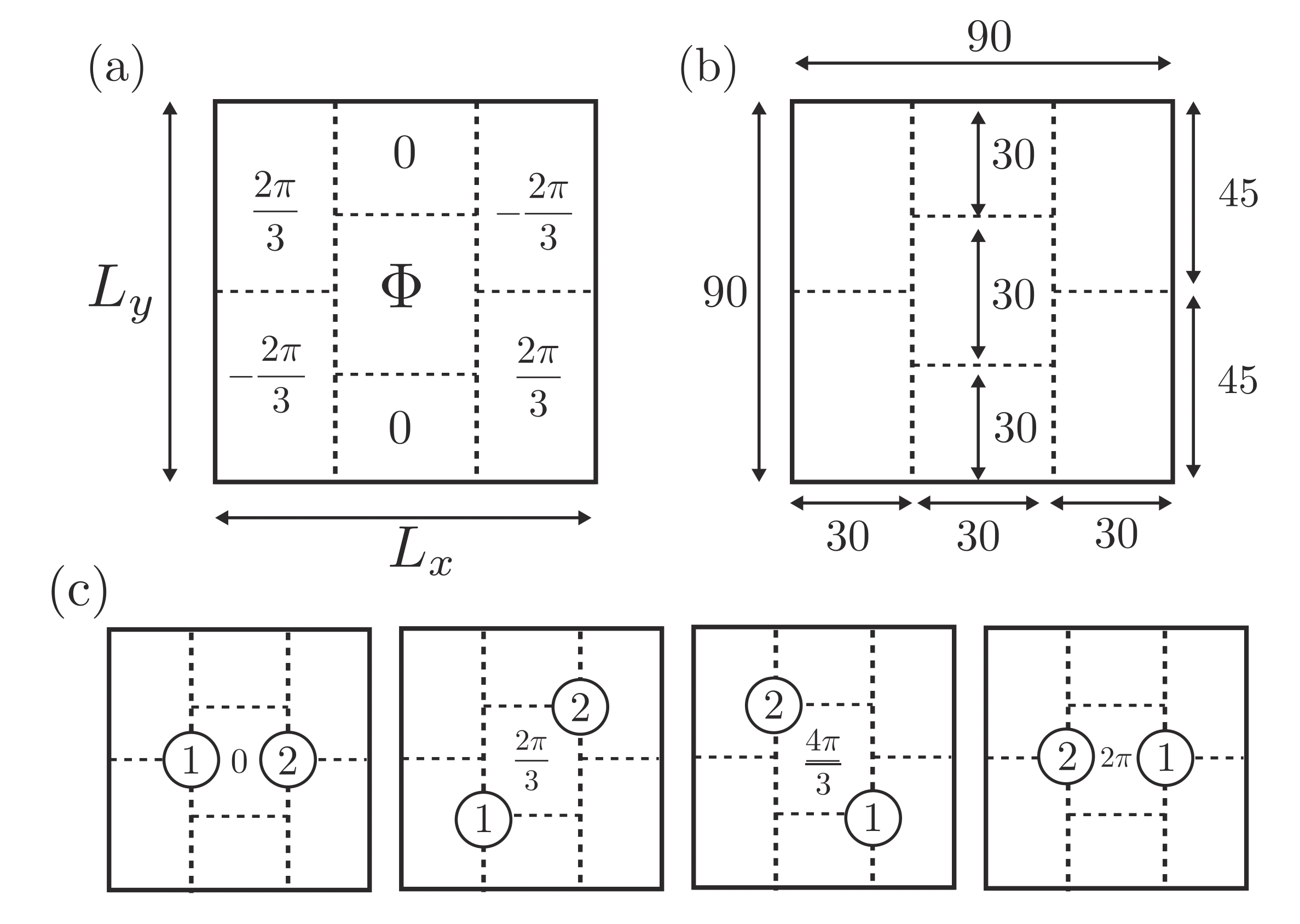}\label{fig:setup_2}} \\
%  \hspace{0.5cm}
  \subfloat{\includegraphics[width=85mm]{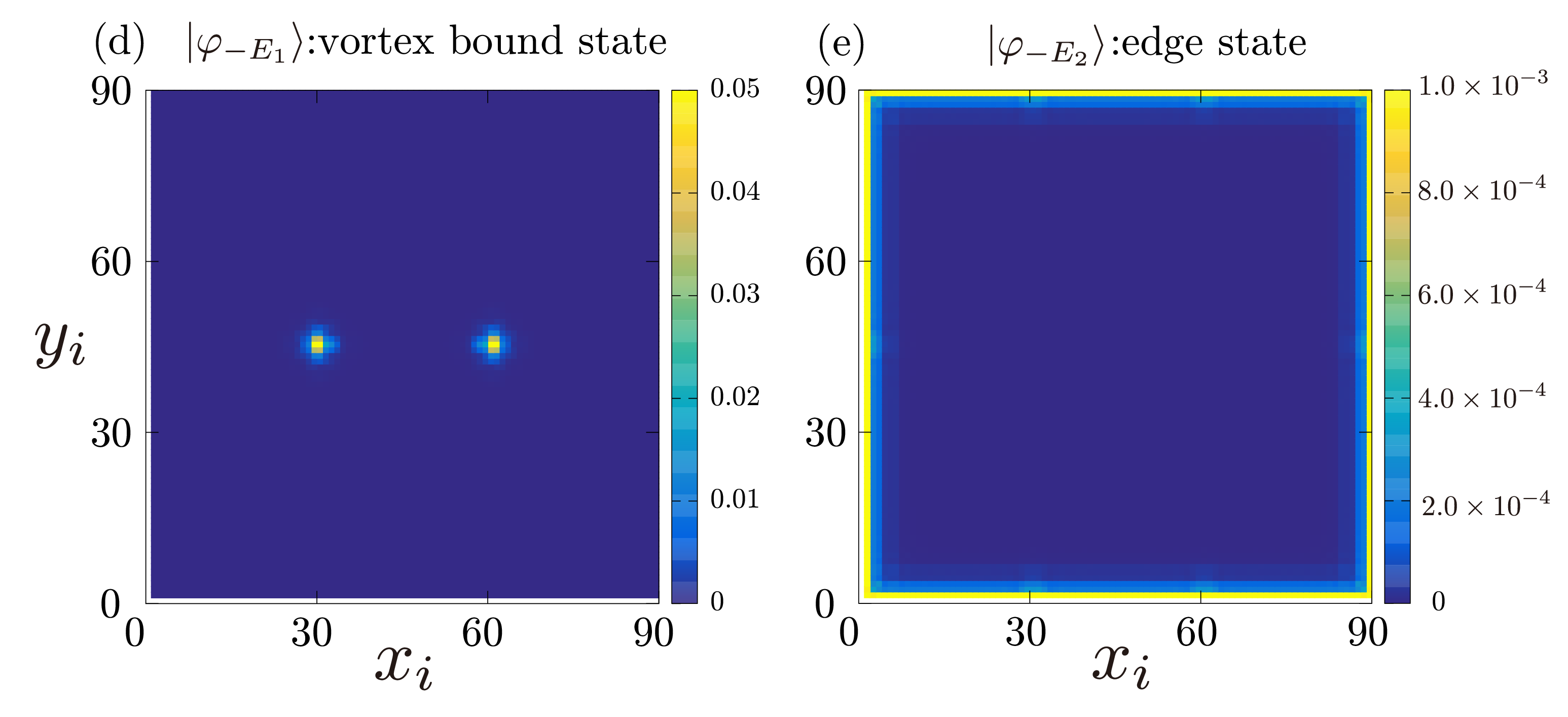}\label{fig:state_2}}
  %\subfloat{\includegraphics[width=5cm,height=5cm]{geometricphase.png}\label{fig:geometricphase}}
  \caption{(a) Schematics of the $U(1)$ phase configuration of the superconducting order parameter for two-vortex simulation. (b) Domain size of calculated systems. (c) Motion of vortex singularities with changing $\Phi$. (d,e) Spatial profiles of quasiparticle wave functions, $\sum _{\sigma}[|u_{{\bm i},\sigma}|^2+|v_{{\bm i},\sigma}|^2]$, in the $\ket{\varphi_{- E_1}}$ and $\ket{\varphi_{- E_2}}$ states when $\Phi=0$. The former (latter) is identified as the vortex (edge) bound state.}
  \label{fig:two}
  %\end{tabular} 
\end{figure}
%=====================================================/

Here we would like to mention two fundamental properties of braiding two vortices in class-D topological superconductors. First, the particle-hole symmetry prohibits transition between particle-hole symmetric MZMs. This can be observed by considering the case of two vortices in Eq.~\eqref{eq:pimatrix}. Owing to the localized nature of the zero energy vortex bound states, the diagonal matrix elements, $\Pi_{nn}$, are exponentially small with respect to $R/\xi$. In the case of two vortices the Berry connection matrix results in 
\begin{equation}
\Pi_{nm}(t) = 0, 
\label{eq:matrix}
\end{equation}
{\it i.e.}, $i\partial_t {C}^{(n)}_m(t)={\bm 0}$. This implies that the fluctuation of the fermion parity $\hat{P}$ is prohibited by the particle-hole symmetry. 
%This result shows that the non-negligible error doesn't happen and says that PHS plays an important role in Majorana condition\eqref{eq:majorana} and Braiding.
%Now, we discuss from the standpoint of BdG theory then must be careful about the space of ground state.

Another fundamental property is that when two vortices are braided, the MZM acquires both the geometric phase and dynamical phase. The former represents the transformation of the quasiparticle basis [the $B$ matrix in Eq.~\eqref{eq:etatn}]. The latter is attributed to the effect of tunneling splitting of MZMs and results in an extra phase, which may generate a non-negligible error during braiding protocols~\cite{chengPRB84}.
According to the braiding rule of strict MZMs, the wavefunction of the MZM acquires the $\pi/2$ phase shift due to the geometric phase, $\braket{\psi(0)|\psi(T)}=e^{\pm i\pi/2}$, which is independent of the detail of braiding operation such as period and trajectories of vortices. This can be obtained by introducing a complex fermion composed of two hybridized MZMs, ($\hat{\gamma}^1_+$, $\hat{\gamma}^2_+$), as $ \hat{\eta}_+ = (\hat{\gamma}^1_+ + i \hat{\gamma}^2_+)/2$. A vortex is accompanied by the branch cut which defines the $2\pi$ phase jump of the order parameter attaching to the vortex singularity. In exchanging two vortices, the MZM moving across the branch cut experiences the phase shift by $\pi$ and the Majorana operators are transformed as $\hat{\gamma}^1_+\rightarrow \hat{\gamma}^2_+$ and $\hat{\gamma}^2_+\rightarrow -\hat{\gamma}^1_+$.
%When the Bogoliubov quasiparticle moves across the branch cut, the phase of the wave function is shifted by $\pi$. When the interchange operation of two vortices transforms the Majorana operators as $\gamma^1_+\rightarrow \gamma^2_+$ and $\gamma^2_+\rightarrow -\gamma^1_+$, 
The complex fermion then changes from $\hat{\eta}_+(0) = (\hat{\gamma}^1_+ + i \hat{\gamma}^2_+)/\sqrt{2}$ to $\hat{\eta}_+(T)=e^{-i\pi/2}\hat{\eta}_+(0)$, which is accompanied by the extra phase factor $e^{-i\pi/2}$. Hence, the accumulation of the $U(1)$ phase after braiding operation,
\beq
|\phi _{\rm geo} (T)| =  \pi /2,
\label{eq:geometric}
\eeq
serves as a direct signature that the braiding dynamics is governed by Eq.~\eqref{eq:umatrix2} with $\vartheta=\pi/4$. The deviation of the accumulated quasiparticle phase from Eq.~\eqref{eq:geometric} implies that $\vartheta(T)$ in the braiding matrix is deviated from $\pi/4$, and the accumulation of extra phases due to the adiabaticity and quasiparticle tunneling gives rise to intrinsic errors in the non-Abelian statistics.

\begin{figure}
%  \centering
  %\begin{tabular}{ccc}
  \subfloat{\includegraphics[width=85mm]{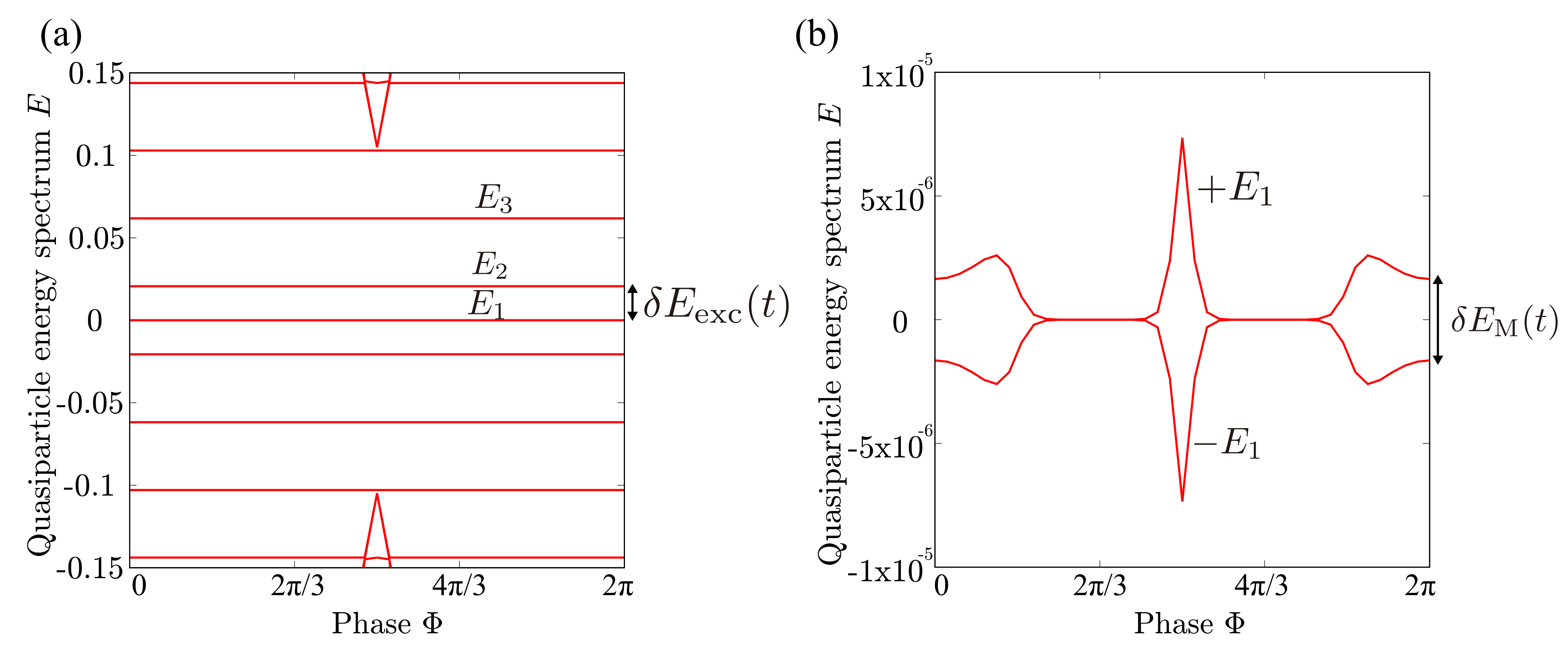}\label{fig:energy_2}}
 \\
  \subfloat{\includegraphics[width=85mm]{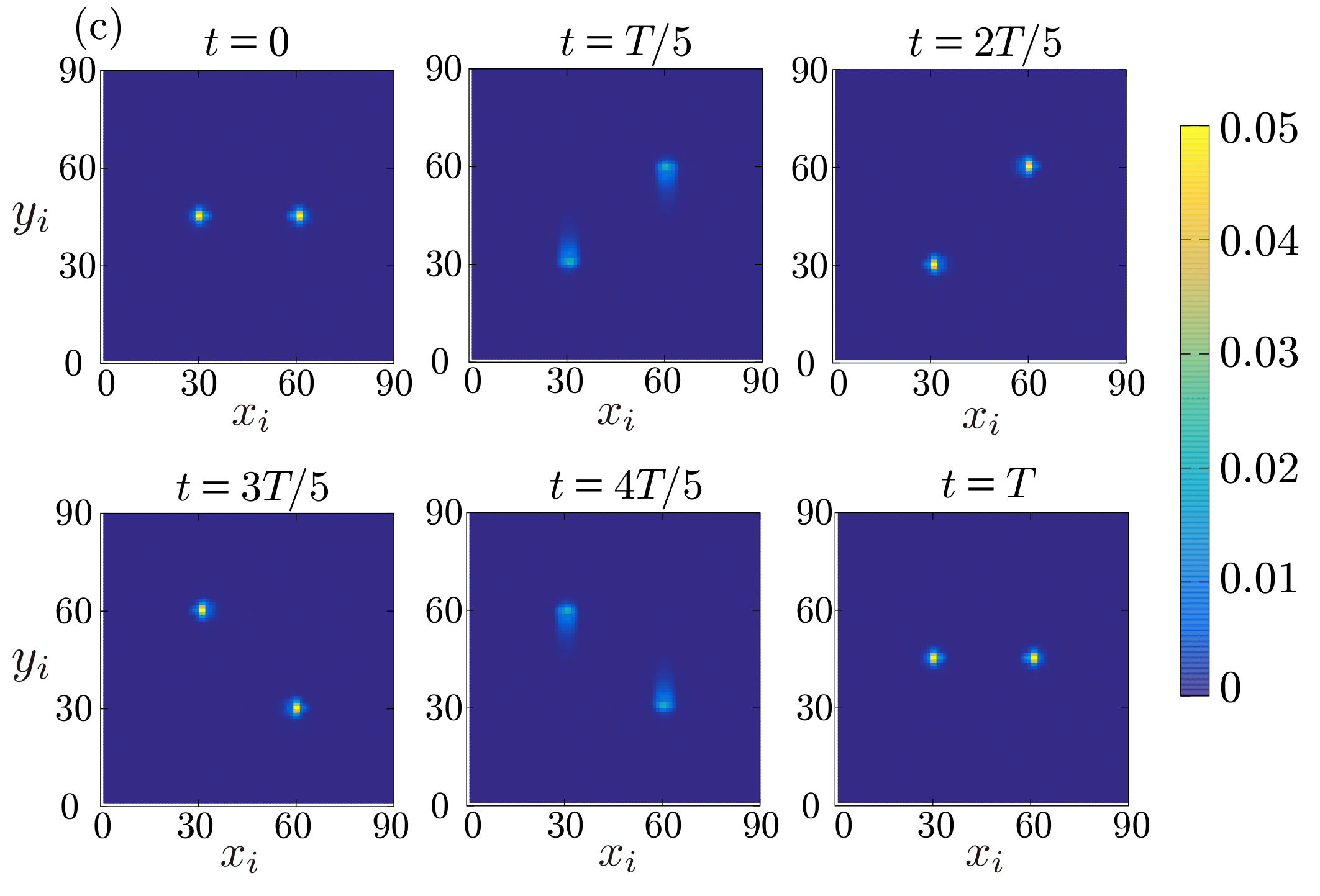}\label{fig:braiding_2}}
  \caption{(a) Evolution of the instantaneous eigenenergies with respect to the phase change of the center domain, $\Phi(t)$. (b) Instantaneous eigenenergies of splitting MZMs, where the maximum splitting occurs at $\Phi=\pi$. (c) Time evolution of the quasiparticle wave functions, $\sum _{\sigma}[|u_{{\bm i},\sigma}(t)|^2+|v_{{\bm i},\sigma}(t)|^2]$, obtained by solving the TDBdG equation \eqref{eq:TBdG}. This can be seen in the Supplemental Material ~\cite{supple}.}
  %\end{tabular} 
 \label{fig:two_result}
\end{figure}

To evaluate the geometric phase $\phi _{\rm geo}$ and the braiding matrix in Eq.~\eqref{eq:umatrix2}, we perform the numerical simulation of the TDBdG equation~\eqref{eq:TBdG}. We also compute Eq.~\eqref{eq:BdGinst} to obtain the instantaneous eigenstates at $t$.  %simulate Majorana braiding dynamics and check $\pi/2$ phase shift. 
We impose the open boundary conditions at the edges of the two-dimensional lattice. Figures~\ref{fig:two}(a) and \ref{fig:two}(b) show the two-dimensional network of the trijunctions, where $\Phi$ denotes the phase of the superconducting order parameter. As shown in Fig.~\ref{fig:two}(c), the braiding operation can be implemented by linearly changing $\Phi$ in the central island from $0$ to $2\pi$, as 
\beq
\Phi(t)= 2\pi t/T.
\eeq
We first prepare the initial state $\ket{\psi_n(t=0)}$ by diagonalizing  $\mathcal{H}_{\rm BdG}$ in Eq.~\eqref{eq:BdGinst} with the order parameter configuration in Fig.~\ref{fig:two}(a) with $\Phi(t=0)=0$. Figures~\ref{fig:two}(d) and \ref{fig:two}(e) show the wave functions of the lowest ($E_1/t_0=1.643 \times 10^{-6}$) and second lowest eigenstates ($E_2/t_0=2.061 \times 10^{-2}$), which are tightly bound at the vortex and edge, respectively. Numerically solving the TDBdG equation~\eqref{eq:TBdG} with the given initial state $\ket{\psi_n(t=0)}$, we simulate the braiding dynamics of two vortices.
The spatial profiles of the quasiparticle wave functions during braiding operation, $\sum _{\sigma}|u_{{\bm i},\sigma}(t)|^2+|v_{{\bm i},\sigma}(t)|^2$, are displayed in Fig.~\ref{fig:two_result}(c), where $\ket{\psi(t)}=[u_{{\bm i},\uparrow}(t),u_{{\bm i},\downarrow}(t),v_{{\bm i},\uparrow}(t),v_{{\bm i},\downarrow}(t)]^{\rm T}$. The two pronounced peaks follow the time evolution of the vortex singularities generated by the evolution of $\Phi(t)$. At $t=T$, the positions of two peaks return to the original positions of vortex singularities, while we will show below that the wave functions acquires an extra phase, and cannot return to the original form. 
In Fig.~\ref{fig:two_result}(a), we plot the low-lying quasiparticle spectrum of the instantaneous $\mathcal{H}_{\rm BdG}[{\bm R}(t)]$ as a function of $t$. The vortex bound states stay around the zero energy in the whole $t$, and the non-zero energy states correspond to the edge bound states. We note that the edges are well spatially separated from the vortex singularities, and the hybridization is negligible during the braiding dynamics. Figure~\ref{fig:two_result}(b) shows the enlarged spectrum within $|E| < 3.0 \times 10^{-6}$ corresponding to the splitting energies $+E_+$ and $-E_+$ induced by quasiparticle tunneling between two vortex singularities.

In the numerical simulation, we take the period $T$ of the braiding protocol to satisfy the adiabatic process in Eq.~\eqref{eq:adiabatic} and the nonadiabatic process in \eqref{eq:period_set} within the splitting MZMs. 
Thus, the braiding period $T$ must satisfy the following condition,
\begin{equation}
    \max \left[ \delta E_{\rm M} (t) \right] \lesssim 1/T \lesssim \min \left[ \delta E_{\rm exc} (t) \right],
    \label{eq:period_condition}
\end{equation}
where  $\delta E_{\rm M} (t) $ is the scales of the splitting energies of MZMs, and $\delta E_{\rm exc} (t)$ is the energy difference between the lowest energy (splitting MZM) state and lowest non-Majorana state (the higher CdGM state or edge state). 
The upper and lower bounds of the braiding period are determined from the instantaneous quasiparticle spectrum.
As seen in Fig.~\ref{fig:two_result}(a), the energy difference between the Majorana band and excited states is almost constant on $t$. As the splitting of MZMs has the maximum value when $ \Phi = \pi $, the upper bound of $T$ is set to the maximum splitting energy. 
Therefore, the condition of $T$ is given by $ 50 \lesssim T \lesssim 6.1 \times 10^5$, and $T$ in the numerical simulation is set to $T=1500t_0^{-1}$, where $dt=0.003 t_0^{-1}$.
We set $\ket{\varphi_{-E_1}({\bm R}(0))}$ as the initial state at $t=0$. The numerical result of the interchange of two vortices shows that  the transition probability from the initial ($-E_1$) state to the particle-hole symmetric ($+E_1$) states is found to be 
\begin{gather}
P^{(-)}_+=|\braket{\varphi_{+E_1}|\psi(T)}|^2 = \mathcal{O}(10^{-16}t^{-1}_0), \\
%\eeq
%and 
P^{(-)}_-=|\braket{\varphi_{-E_1}|\psi(T)}|^2 = 0.997 .%\approx 1
\end{gather}
%In Fig.~\ref{fig:two_time}(a), we summarize the transition probabilities $P^{(-)}_{\pm}$ as a function of the braiding period $T$. 
This result is consistent with the transition rule in Eq.~\eqref{eq:pimatrix}, where the direct transition between splitting MZMs in $N=1$ is protected by the particle-hole symmetry.

%---------------------------------------------------
%two vortices results figures
%---------------------------------------------------

\begin{figure}[t!]
%\centering
\includegraphics[width=85mm]{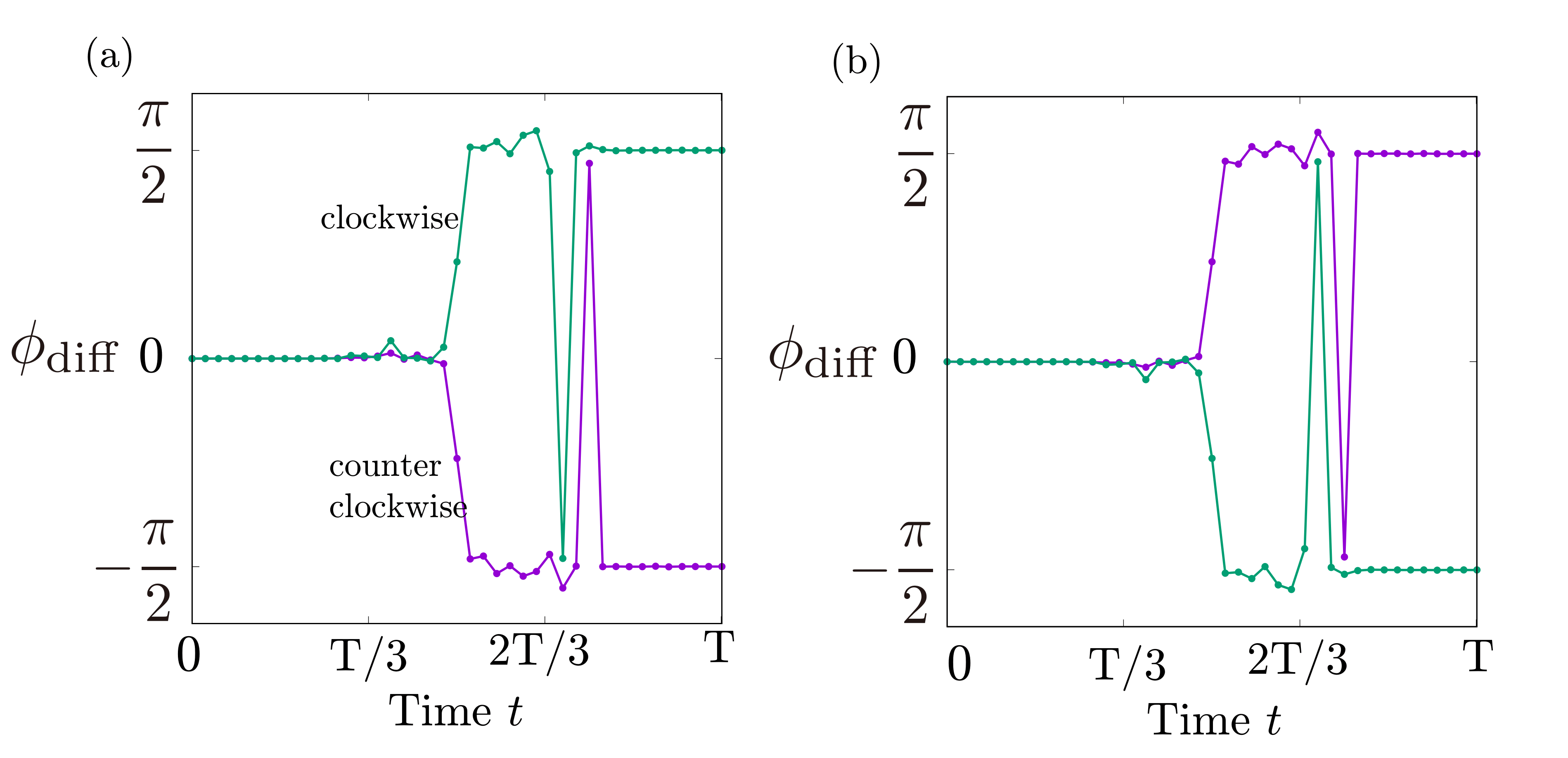}
            \caption{Time evolution of the phase difference $\psi_{\rm diff}(t)$ in the $\ket{\varphi_{-E_1}}$ state (a) and the $\ket{\varphi_{+E_1}}$ state (b) in the clockwise (green) and counterclockwise (purple) interchange operation of two vortices. }
\label{fig:two_phase}
\end{figure}

We now extract the dynamical and geometric phases from the numerical simulation of braiding vortices.
The accumulation of the phase in braiding dynamics is given as 
\begin{equation}
    \phi_{\rm diff} (t)= \arctan \left\{\frac{{\rm Im} \braket{\psi(0)|\psi(t)}}{{\rm Re} \braket{\psi(0)|\psi(t)}}\right\}.
\end{equation}
Figures \ref{fig:two_phase}(a) and \ref{fig:two_phase}(b) show the time evolution of the phase difference $\phi_{\rm diff}(t)$ in $\ket{\varphi_{-E_1}}$ and $\ket{\varphi_{+E_1}}$, respectively. Here we plot $\phi _{\rm diff}(t)$ for the counterclockwise (purple) and clockwise (green) rotation of vortices, which can be implemented by changing the phase $\Phi(0) = 0 \rightarrow \Phi (T) = 2 \pi$ and $\Phi(0)=0 \rightarrow \Phi(T)=-2 \pi$ in Fig.~\ref{fig:two}(a), respectively. 
We find that after interchange operation the vortex bound states accumulate the phase $|\phi_{\rm diff}(T)| =1.572\approx \pi /2$, which contains the contributions of both the geometric phase and the dynamical phase. From the energy spectrum and braiding period $T = 1500t_0^{-1}$, the contribution of the dynamic phase factor to $\phi_{\rm diff}(T)$ is estimated as $\mathcal{O}(10^{-3})$. As $T$ increases, the braiding dynamics approaches the adiabatic regime, $T\gg \delta E^{-1}_{\rm M}$, and the noise effect, $\phi _{\rm diff}(T)-\pi/2$, is induced by the dynamical phase. In Fig.~\ref{fig:two_phase}, it is seen that $\phi _{\rm diff}(t)$ abruptly jumps around $t=3T/2$. This is attributed to the peculiarity of the trijunction model in Fig.~\ref{fig:two}(a), where the signs of vorticities of both vortex singularities after the period are inverted from those in the initial state. However, it turns out that the sign flip of vorticity does not affect the phase accumulation of vortex bound states.

%======================================
%=====================================/

The counterclockwise motion of vortices is the time reversal symmetric operation of the clockwise one. The quasiparticles along the two trajectories acquires geometric phase with opposite sign while accumulating the same dynamical phase.
To extract the geometric phase from $\phi _{\rm diff}$, therefore, we decompose $\phi^{\rm clock}_{\rm diff}(T)$ and $\phi^{\rm count}_{\rm diff}(T)$ accumulated by the clockwise and counterclockwise interchange as
\begin{gather}
  \phi^{\rm clock}_{\rm diff}(T) =  \phi_{\rm geo} + \phi_{\rm dyn}, \\
  \phi^{\rm count}_{\rm diff}(T) = -\phi_{\rm geo} + \phi_{\rm dyn} .
\end{gather}
The numerical simulation of braiding two vortices shows that the accumulated geometric phase coincides with the expected value $\pi/2$ within numerical accuracy, 
\beq
\phi _{\rm geo} = [\phi^{\rm clock}_{\rm diff}(T)-\phi^{\rm count}_{\rm diff}(T)]/2 = -1.57079.
\eeq
Hence, the deviation of the accumulated phase from $\pi/2$ is attributed to the dynamical phase stemming from the hybridization of MZMs. 

After the interchange of two vortices, the lowest vortex-bound quasiparticle state accumulates the geometric phase $\phi _{\rm geo} \approx \pi/2$ and the dynamical phase, $\hat{\eta}_+(T) = ie^{i\delta \phi }\hat{\eta}_+(0)$. The deviation of the accumulated phase, $\delta \phi \equiv \phi _{\rm diff}(T)-\pi/2 \approx \phi _{\rm dyn}$, is attributed to the dynamical phase due to the interaction of MZMs. The interchange of vortices generates the transformation of the Majorana operators, $\hat{\gamma}^{1}_+(t)\equiv[\hat{\eta}_+(t)+\hat{\eta}^{\dag}_+(t)]/\sqrt{2}$ and $\hat{\gamma}^{2}_+(t)\equiv -i [\hat{\eta}_+(t)-\hat{\eta}^{\dag}_+(t)]/\sqrt{2}$, as 
\begin{gather}
\hat{\gamma}^{1}_+(T)=-\cos(\delta\phi)\hat{\gamma}^{2}_+(0)-\sin(\delta\phi) \hat{\gamma}^{1}_+(0)  \\
\hat{\gamma}^{2}_+(T)=\cos(\delta\phi) \hat{\gamma}^{1}_+(0) -\sin(\delta\phi)\hat{\gamma}^{2}_+(0) .
\end{gather}
This braiding dynamics of quasiparticles is represented by the braiding matrix $U(T)$ in Eq.~\eqref{eq:umatrix2} with 
\beq
\vartheta (T) = \frac{\pi}{4} - \delta \phi.
\eeq
Hence, $\delta \phi$ is a possible source of quantum disturbance of non-Abelian braiding dynamics.

\begin{figure}
%\centering
\subfloat{\includegraphics[height=40mm]{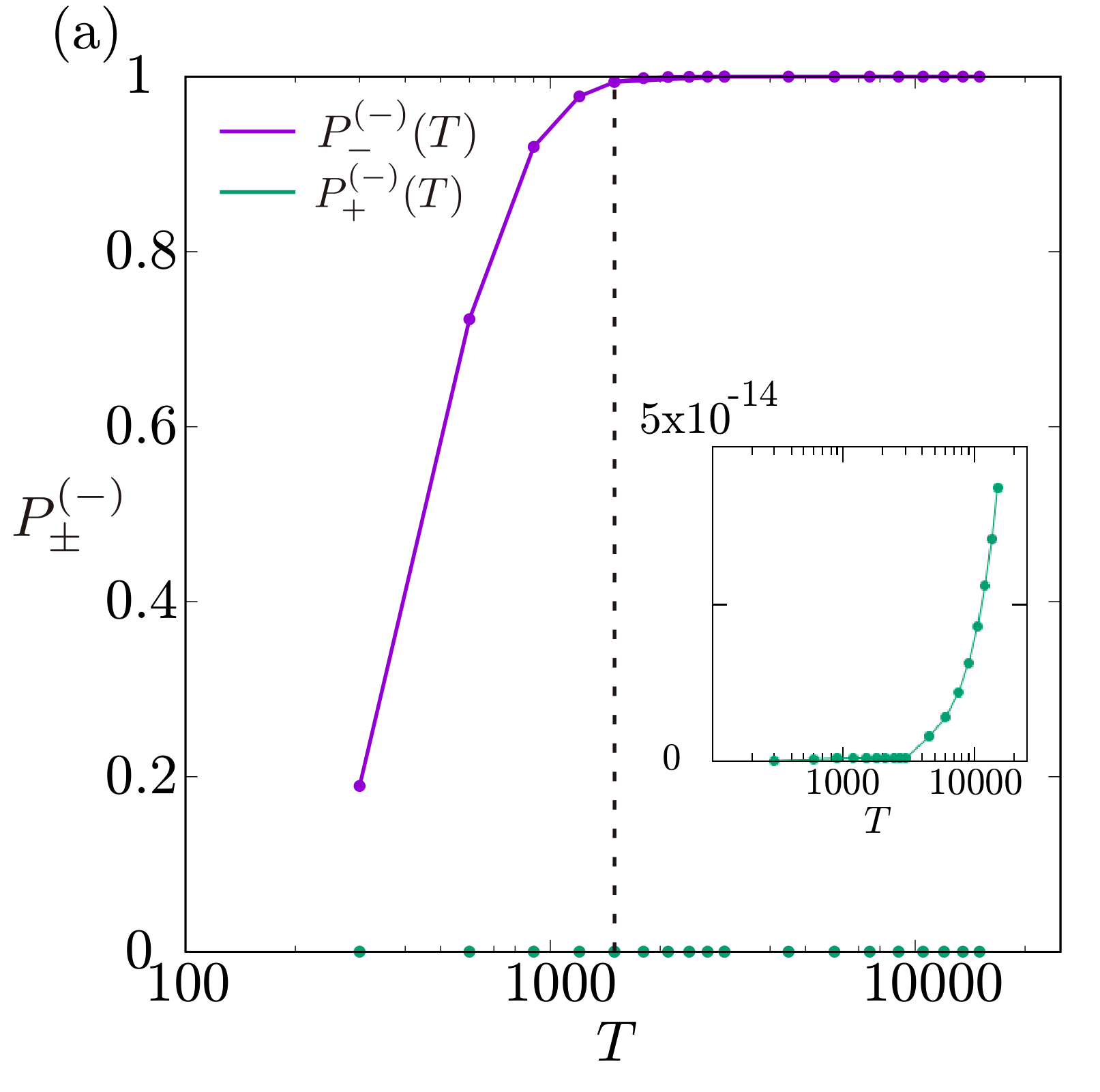}\label{fig:transit_time}}
\subfloat{\includegraphics[height=40mm]{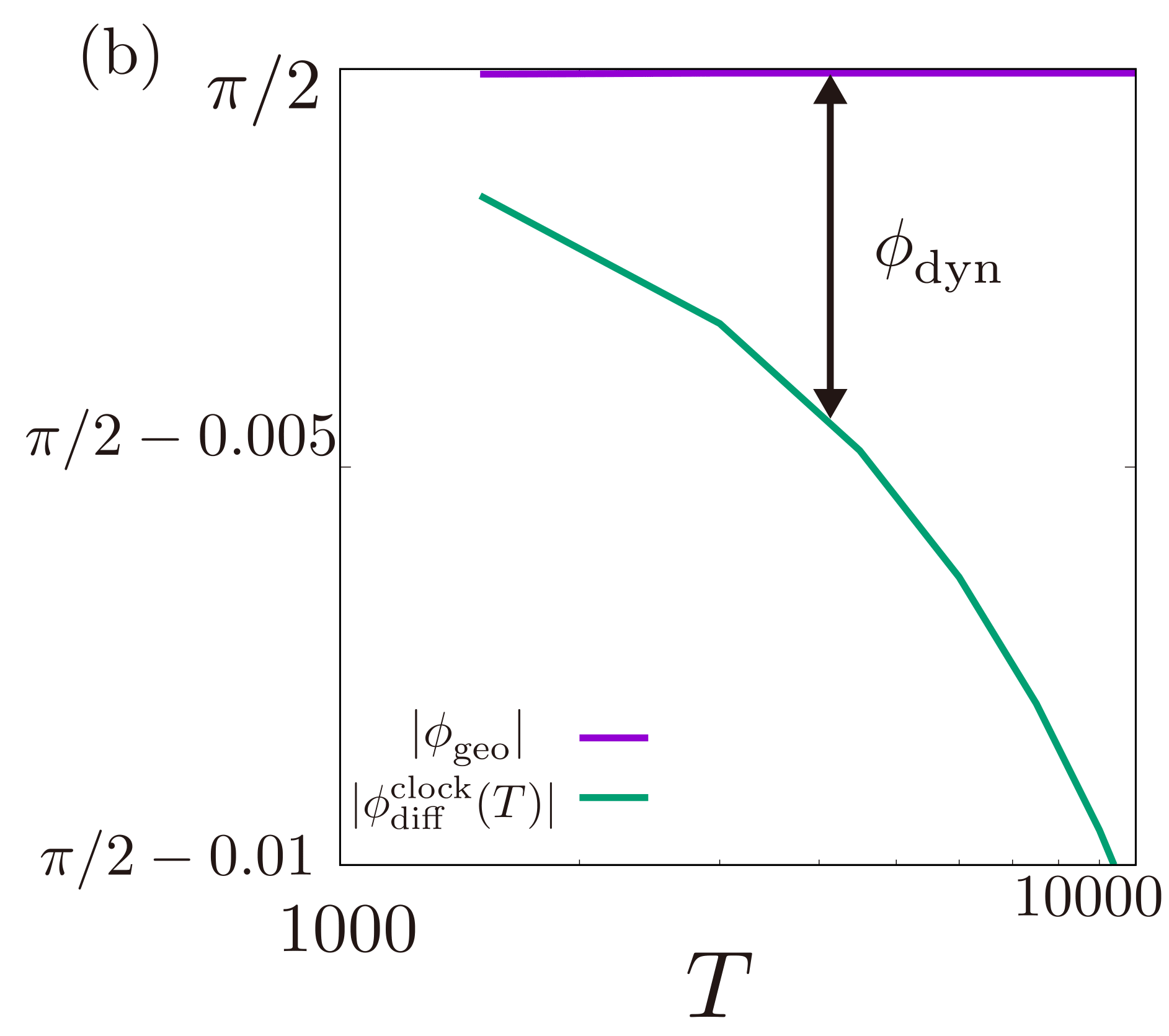}\label{fig:phase_time}}
\caption{Transition probabilities $P^{(-)}_{\pm}$ (a) and the geometric phase $\phi _{\rm geo}$ (b) as a function of the period of the braiding operation, $T$. In Sec.~III, we have discussed the numerical simulation of braiding vortices with $T=1500t^{-1}_0$ which is denoted by the dashed line in (a).
}
\label{fig:two_time}
\end{figure}

In Fig.~\ref{fig:two_time}, we compute the transition probability $P^{(-)}_{\pm}$ and the noise of the accumulated phase $\delta \phi$ as a function of the braiding period $T$. We find that the transition probabilities are $P^{(-)}_{-}\approx 1$ and $P^{(-)}_{+}\approx 0$, and the the noise of the accumulated phase is negligible, $\delta\phi \ll 1$, when the braiding period satisfies $ 1000t^{-1}_0 \lesssim T \
 10000t^{-1}_0$. 
For $T\lesssim 1000t^{-1}_0$, however, $P^{(-)}_-$ significantly decreases with increasing the speed of the braiding operation. The depletion of the norm $P^{(-)}_-\equiv|\braket{\varphi_{-E_1}|\psi(t)}|^2$ is attributed to the nonadiabatic transition to the higher CdGM states. 
These results numerically demonstrate that when the braiding operation satisfies the conditions in Eqs.~\eqref{eq:adiabatic} and \eqref{eq:period_set}, the noise effect due to the interaction of neighboring MZMs is negligible, and the non-Abelian braiding dynamics can be successfully accomplished as in Eq.~\eqref{eq:umatrix2} with $\vartheta (T) \approx \pi/4$. 
We also note that for $T\gtrsim 10000t^{-1}_0$, the transition probability between the particle-hole symmetric states, $P^{(-)}_+\equiv |\braket{\varphi_{+E_1}(0)|\psi(t)}|^2$, exponentially increases with respect to $T$. This corresponds to the adiabatic regime that the braiding period is slower than the time scale of the energy splitting of MZMs, $T \gg \delta E^{-1}_{\rm M}\sim \mathcal{O}(10^{5}t^{-1}_0)$. As the adiabatic regime is approached, the braiding dynamics of quasiparticles significantly accumulates the dynamical phase, which gives rise to serious errors in the braiding matrix $U(T)$ from $\vartheta(T)=\pi/4$.
%======================================
%=====================================/

\section{Braiding dynamics in four-vortex systems}
\label{sec:four}

In this section, we consider the braiding dynamics of systems with four vortices. 
Let us first consider an ideal situation that vortices are well separated from each other, and each vortex hosts a single MZM. Let $\hat{\gamma}^i$ be the Majorana operator bound at the $i$th vortex ($i=1,2,3,4$), and $\hat{c}_{12}=(\hat{\gamma}^1+i\hat{\gamma}^2)/\sqrt{2}$ and $\hat{c}_{34}=(\hat{\gamma}^3+i\hat{\gamma}^4)/\sqrt{2}$ be the operators of complex fermions. The $2$-dimensional Fock space is spanned by the degenerate ground states, $\ket{00}$ and $\ket{11}=\hat{c}^{\dag}_{12}\hat{c}^{\dag}_{34}\ket{00}$, which defines a single qubit. The manipulation of the qubit can be implemented by the interchange of $i$th and $j$th vortices, $U_{ij}$, which is defined in Eqs.~\eqref{eq:braidingU} and \eqref{eq:umatrix2}. The operation, $U_{12}(T)$, leads to the $\pi/4$ phase rotation of the qubit, while the  operations, $U_{13}(T)$ and $U_{13} (2T)=[U_{13} (T)]^2$ implement the Hadamard gate, $\ket{00} \rightarrow ( \ket{00} + \ket{11} )/ \sqrt{2}$, and the NOT gate, $\ket{00}\rightarrow \ket{11}$, respectively. 
%======================================
\begin{figure}
%  \centering
  \subfloat{\includegraphics[width=85mm]{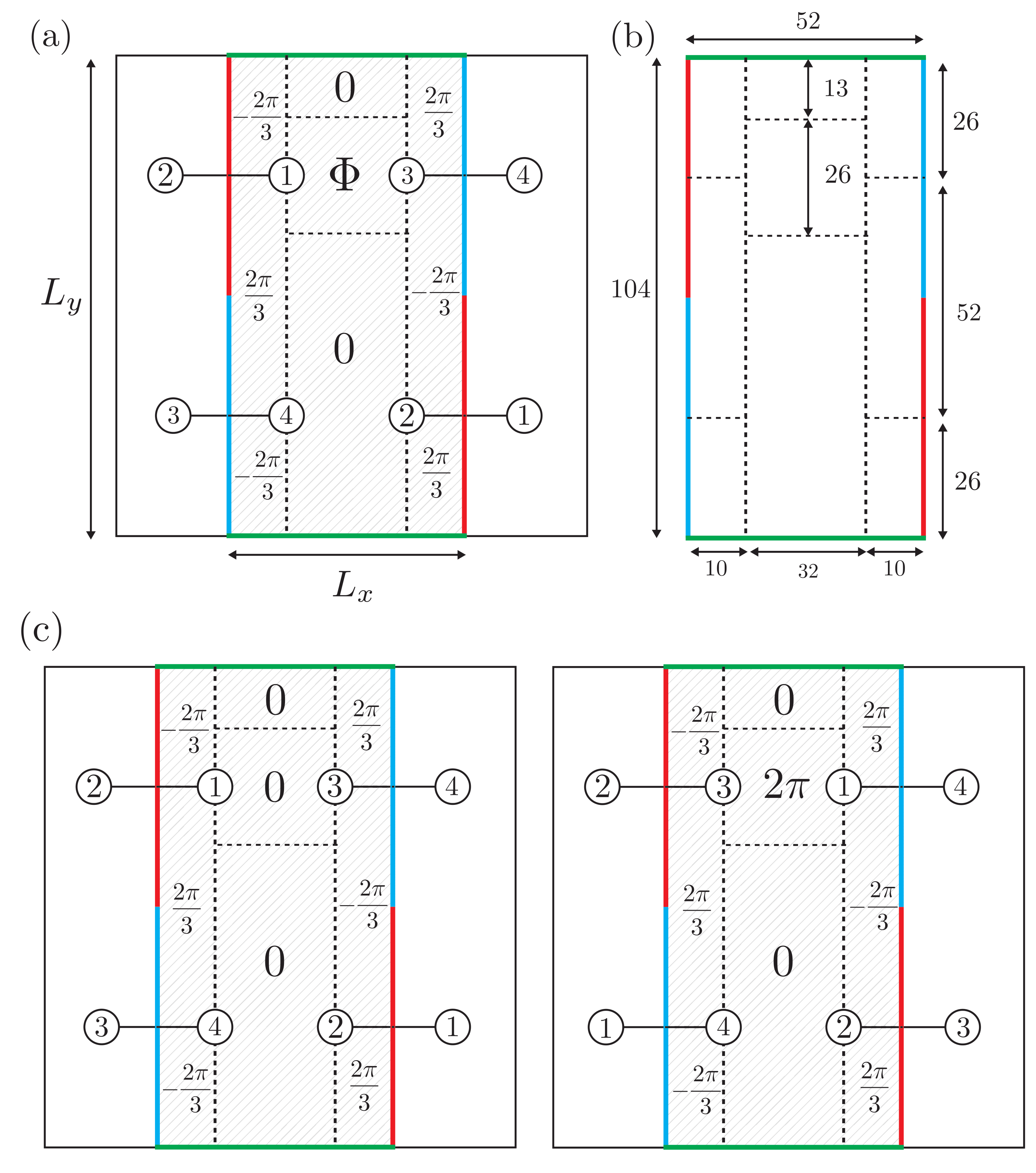}}
  \\ \relax
	\subfloat{\includegraphics[width=85mm]{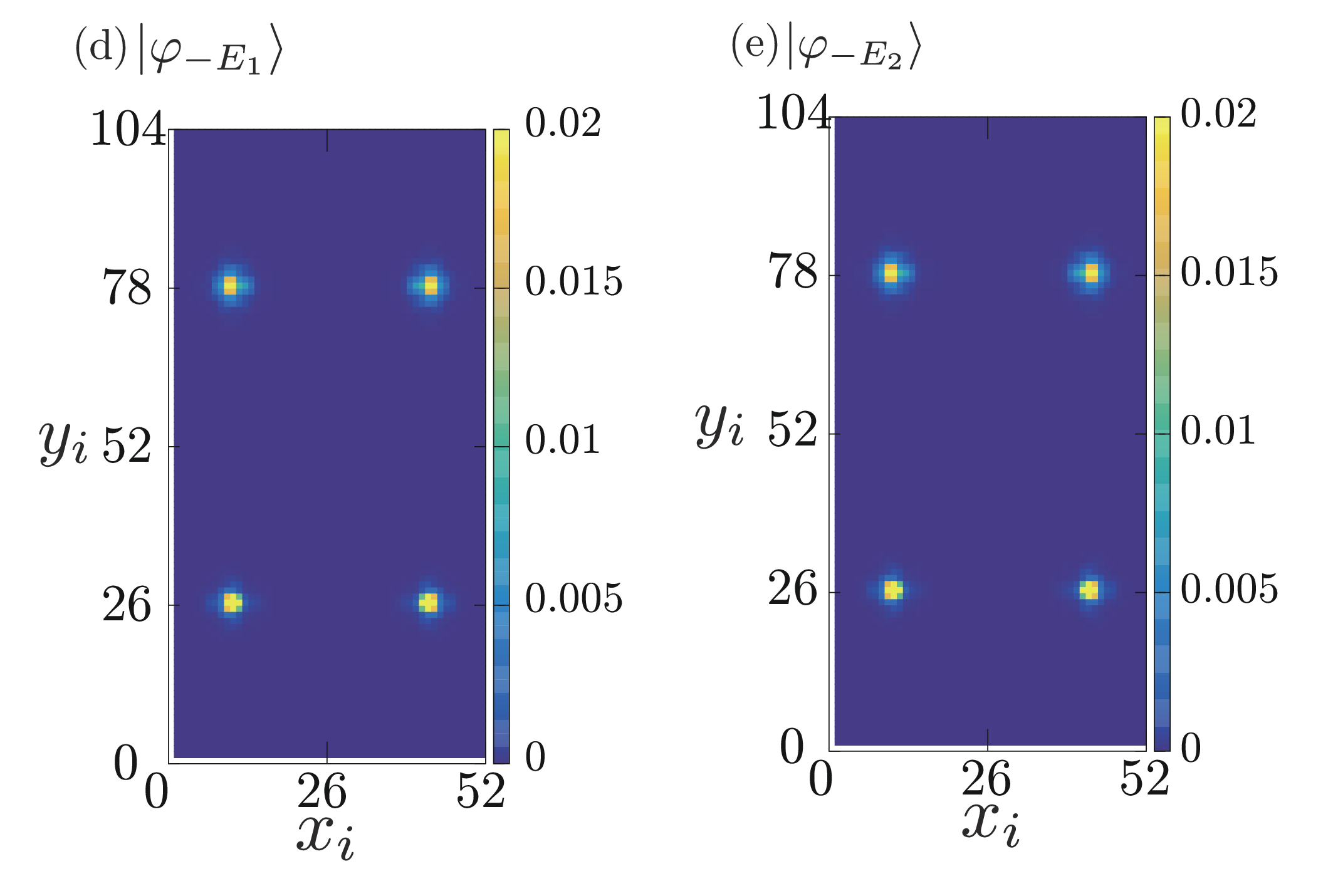}}
  \caption{(a,b) Configuration of four vortices made from superconducting junctions for numerical simulation of braiding dynamics. The lines with same color are identified by the periodic boundary conditions in Eqs.~\eqref{eq:pbc1} and \eqref{eq:pbc2}. The dash region is the unit cell of tri-junction network and its numerical set up is fig.(b) satisfies the condition $D_{12}, D_{34} < D_{13}, D_{24}$. (c) The advance of $\Phi$ in a superconducting domain interchanges vortex singularities labeled by ``1'' and ``3''. (d,e) Spatial profiles of the quasiparticle wave functions with the energy $-E_1$ and $-E_2$, $\sum _{\sigma}[|u_{{\bm i},\sigma}|^2+|v_{{\bm i},\sigma}|^2]$, where the $U(1)$ phase is set to be $\Phi=-2\pi$.}
  \label{four_phase}
\end{figure}
%=====================================/

Here we examine the noise effect on non-Abelian statistics of vortices and quantum gates by numerically simulating the TDBdG equation with four vortices. In this section we present the numerical simulation in the junctions of $s$-wave superconductors with the Rashba SOI. The braiding dynamics in the Fu-Kane model, which is the heterostructure of a topological insulator and an $s$-wave superconductor, is shown in Appendix A.
As the zero energy vortex-bound states in class D topological superconductors are protected by a $\mathbb{Z}_2$ invariant, and the quasiparticle tunneling between neighboring vortices during braiding operation causes the non-negligible splitting of the ground states. To address the effect of quasiparticle hybridization, we consider the four-vortex system shown in Fig.~\ref{four_phase}(a), which is composed of a two-dimensional array of the superconducting domains with different $U(1)$ phases. To exclude the contribution of edge states, we impose the periodic boundary condition. The size of the unit cell is $L_x\times L_y$ and the periodic boundary conditions along the $x$- and $y$-directions are imposed as 
\begin{gather}
\hat{\Delta}(i_x=L_x,i_y) = \hat{\Delta}(i_x=1,i_y=i_y+L_y/2), 
\label{eq:pbc1} \\
\hat{\Delta}(i_x,i_y=L_y) = \hat{\Delta}(i_x,i_y=1),\label{eq:pbc2}
\end{gather}
respectively. Below we mainly show the numerical results for $L_x = 52$ and $L_y=104$ [see Fig.~\ref{four_phase}(b)]. As shown in Fig.~\ref{four_phase}(c), the phase configuration at $\Phi=0$ hosts the four vortex singularities labeled by ``1'', ``2'', ``3'', and ``4''. 
Without loss of generality, we take the condition $D_{12}, D_{34}< D_{13},D_{24}$, where 1-2 and 3-4 vortices are tightly paired. In the numerical calculation, we set $D_{12} = D_{34} = 20$ and $D_{13}=D_{24}=32$, where $D_{12}$ and $D_{34}$ ($D_{13}$ and $D_{24}$) are the intra-pair (inter-pair) distance. Without loss of generality, we take the condition $D_{14}, D_{23}< D_{13},D_{24}$. The set of other parameters is same as that in Sec.~\ref{sec:two}.
The rotation of the superconducting phase $\Phi$ interchanges the inter-pair vortex singularities labeled by ``1'' and ``3'' in a counterclockwise direction.
We start with $ \Phi = -2 \pi $ and evolve $\Phi$ to $2\pi$ to implement the twice interchange of ``1'' and ``3'' vortex singularities. 
%The spatial profiles of the quasiparticle wave function for four-vortex system are displayed in Fig.\ref{four_phase}(d)(e).
%It says that a vortex bound state is obtained at the position expected from the phase structure (Fig.\ref{four_phase}(c)).
%We set $\ket{\varphi_{-E_1}}, \ket{\varphi_{-E_2}} $ as the initial states.

%Parameter setting is same in Sec.\ref{sec:two} and we set the phase geometry as Fig.\ref{four_phase}(a) where color line means periodic boundary condition (b).
%This boundary condition exclude the contribution of edge states.
%It is expected that the contribution of edge states can be neglected by setting the system size large enough.
%Generally, the main contribution is considered the effect of mini gap $\Delta^2/E_F$ (CdGM states) in vortex bound states. 
%Fig.\ref{four_phase}(c) is the expected behavior of vortices in numerical calculations.

%======================================
\begin{figure}
%  \centering
  \subfloat{\includegraphics[width=85mm]{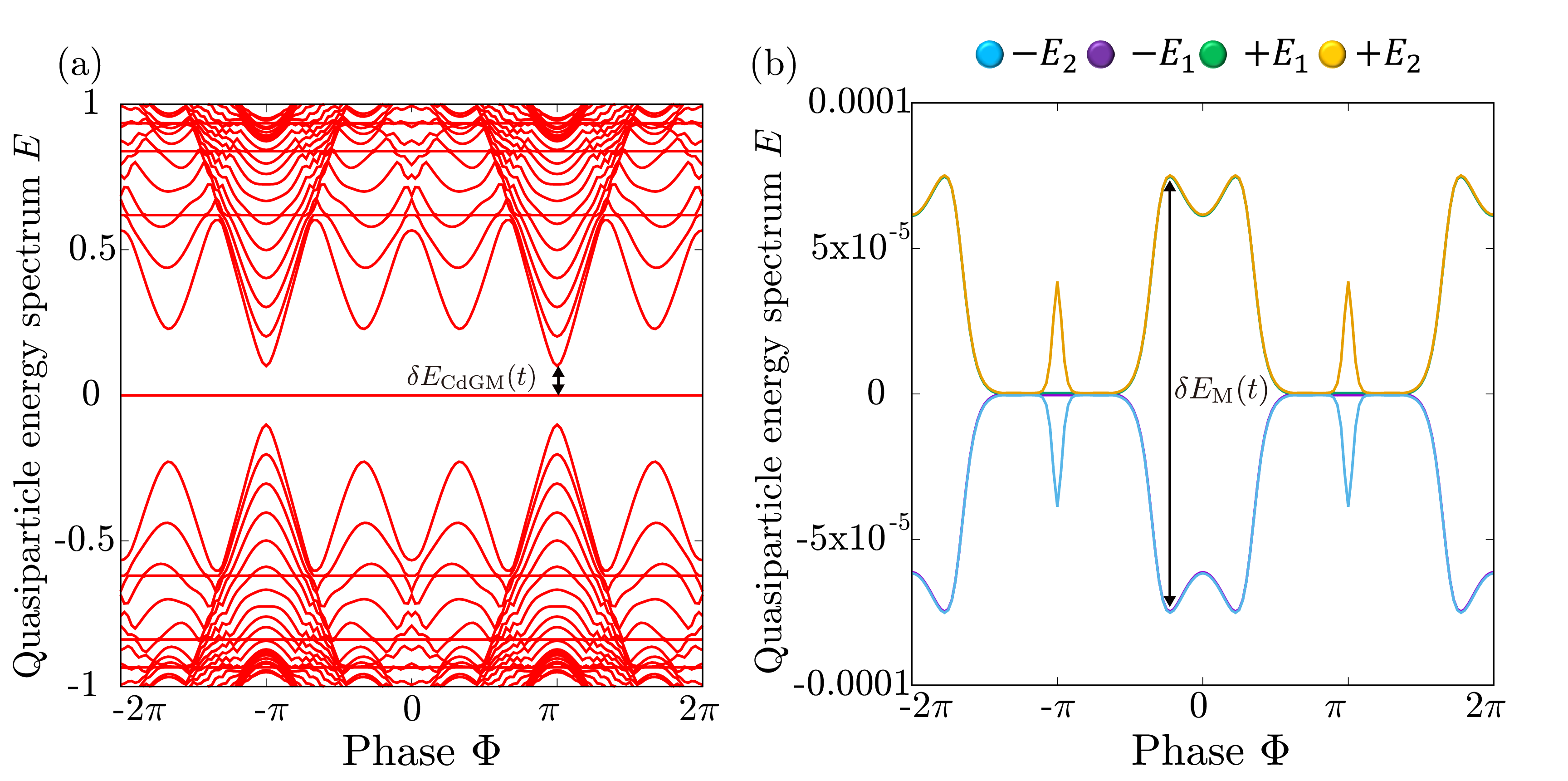}
  \label{fig:energy_4}}\\
  \subfloat{\includegraphics[width=85mm]{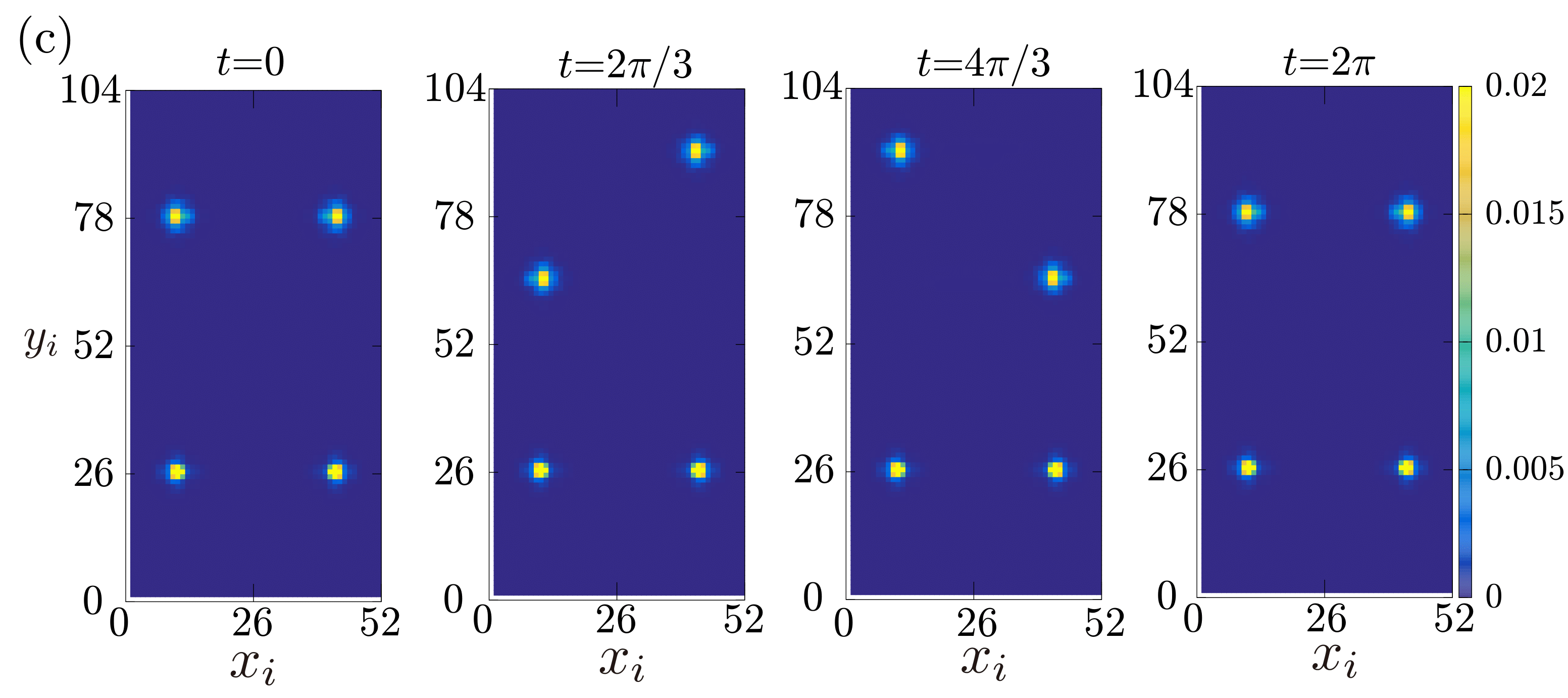}
  \label{fig:braid_4}}
%  \\ \relax
  \caption{(a) Evolution of the instantaneous eigenenergies with respect to the $U(1)$ phase $\Phi$. The lower bound of the braiding period $T$ is determined by the inverse of the minimum splitting between MZMs and the higher CdGM states around $\Phi=\pm \pi$, $\delta E^{-1}_{\rm CdGM}(t)$. (b) Instantaneous eigenenergies of splitting MZMs (Majorana band), where the inverse of the maximum splitting occurs around $\Phi=\pm \pi/150$, $\delta E^{-1}_{\rm M}(t)$, determines the upper bound of $T$. (c) Time evolution of the quasiparticle wave functions, $\sum _{\sigma}[|u_{{\bm i},\sigma}(t)|^2+|v_{{\bm i},\sigma}(t)|^2]$, is obtained by solving the TDBdG equation \eqref{eq:TBdG}. This can be seen in the Supplemental Material ~\cite{supple}.}
  \label{four}
\end{figure}
%=====================================/

When the intervortex distance is macroscopically large, the ground states are doubly degenerate and the braiding of inter-pair vortices can manipulate the degenerate ground states. In numerical calculation, however, the finite size effect gives rise to the quasiparticle tunneling between neighboring vortices and the hybridization of MZMs hosted by each vortex. The splitting energy levels of four MZMs are referred to as $\pm E_1$ and $\pm E_2$, where $E_1<E_2$. Figures~\ref{four_phase}(d) and (e) show the quasiparticle wave functions of the lowest ($E_1/t_0 = 6.118 \times 10^{-5}$) and second lowest $(E_2/t_0 = 6.155 \times 10^{-5})$, which are tightly bound at vortex singularities. Figure~\ref{four}(a) also shows the instantaneous eigenenergies of the BdG Hamiltonian with varying $\Phi$ from $-2\pi$ to $2\pi$. When $\Phi = \pm \pi$, the energy level spacing between the Majorana band and higher CdGM states, $\delta E_{\rm CdGM}(t)$, has the minimum value, which defines the lower bound of the braiding period $T$. Figure~\ref{four}(b) shows the quasiparticle energy spectrum around zero energy, i.e., the Majorana band. The energy width of the Majorana band, $\delta E_{\rm M}(t)$, becomes maximum at $\Phi=\pm \pi/150$. This determines the upper bound of $T$. Hence, the condition of the braiding period $T$ is evaluated as $1.77t^{-1}_0<T<2.7 \times 10^6t^{-1}_0$.
In numerical simulation of the TDBdG equation \eqref{eq:TBdG}, we take the braiding period as $T=1440t^{-1}_0$ and $dt=0.003t^{-1}_0$.

%======================================
\begin{figure}
%  \centering
  \subfloat{\includegraphics[width=85mm]{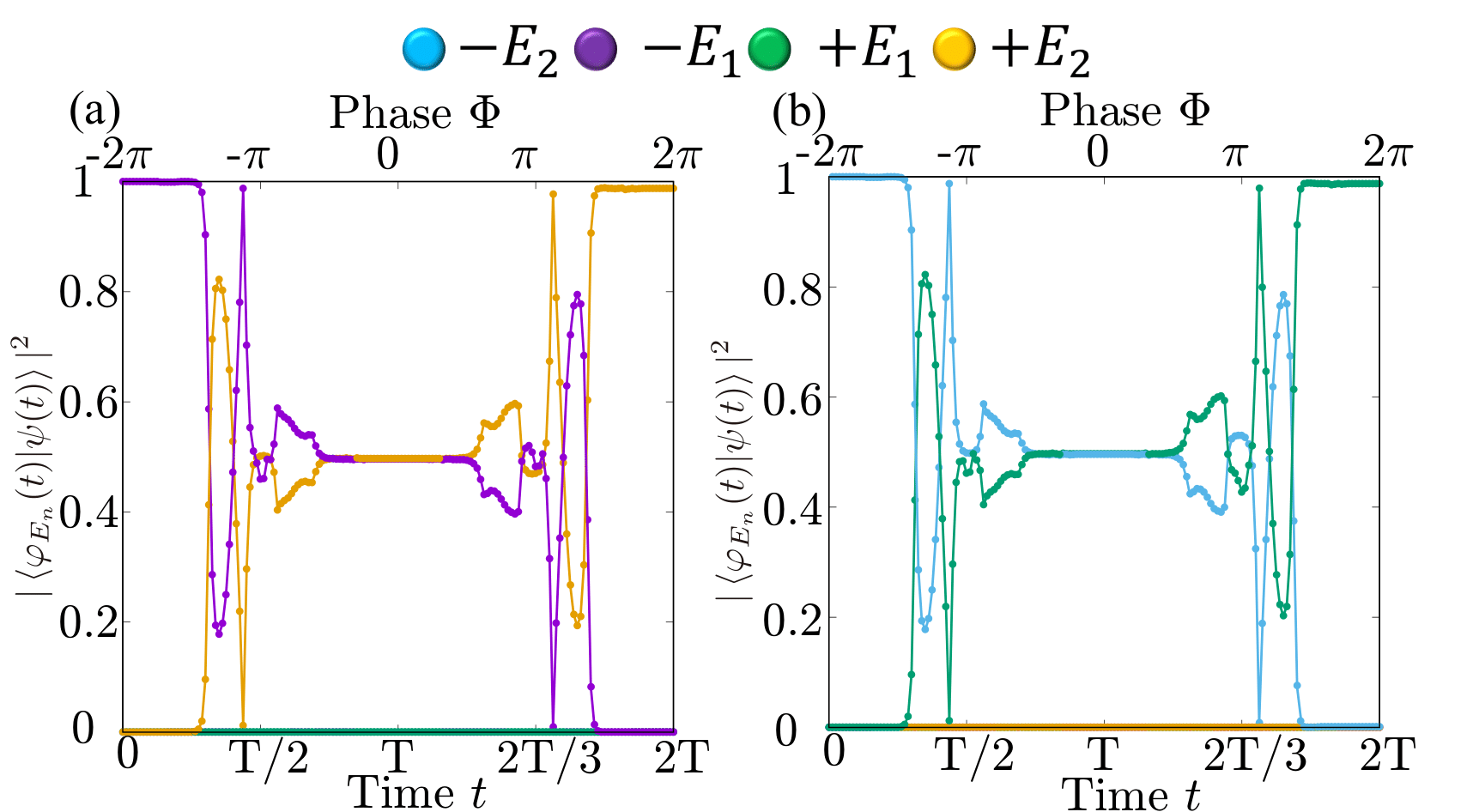}
  \label{fig:project_4}} \\
  \subfloat{\includegraphics[width=85mm]{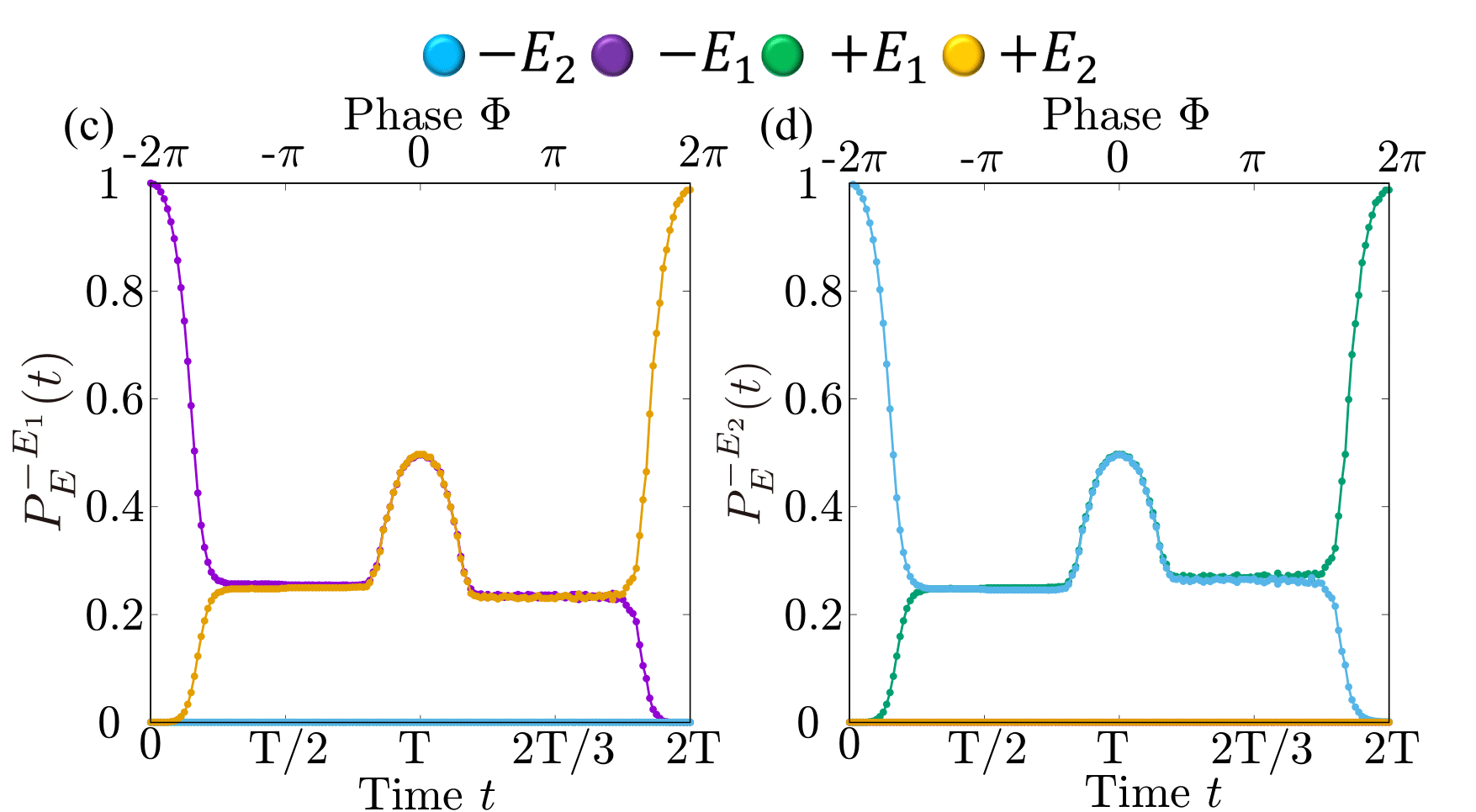}
  \label{fig:transit_4}}
\caption{(a,b) Projections of $\ket{\psi (t)}$ onto the instantaneous eigenstates $\ket{\varphi_{E_n} (\Phi(t))}$, where the initial states in (a) and (b) are $\ket{\varphi_{-E_1} (\Phi(0))}$ and $\ket{\varphi_{-E_2} (\Phi(0))}$, respectively. (c,d) Projection of $\ket{\psi (t)}$ onto the encoded state $\ket{\varphi_{-E_1} (\Phi(0))}$ (c) and $\ket{\varphi_{-E_2} (\Phi(0))}$ (d). The twice interchange of vortices ($t=2T$) causes the transition of the quasiparticle states with $-E_1$ and $-E_2$ to the eigenstates with $+E_2$ and $+E_1$, respectively. This transition implies the unitary transformation of the encoded state $\ket{00}$ to $\ket{11}\equiv \hat{\eta}^{\dag}_{E_1}\hat{\eta}^{\dag}_{E_2}\ket{00}$. }
\label{fig:four_result}
\end{figure}
%=====================================/

We now present the numerical simulation of the braiding dynamics of $\hat{\eta}^{\dag}_{E_{-1}}\hat{\eta}_{E_{-2}}$. We compute the TDBdG equation \eqref{eq:TBdG}, with the initial conditions, $\ket{\psi(t=0)}= \ket{\varphi_{E_{-1}}}$ and $\ket{\psi(t=0)}=\ket{\varphi_{E_{-2}}}$. The $U(1)$ phase is evolved from $\Phi(t=0)=-2\pi$ to $\Phi(t=2T)= 2\pi$ linearly on time, $\Phi(t) = -2\pi + 2\pi t /T$. Figure \ref{four}(c) shows the time-evolution of the quasiparticle wave functions from $t =0$ to $t=T$. The peaks of the quasiparticle wave functions trace the motion of vortex singularities driven by the ${\rm U}(1)$ phase rotation. The numerical results thus demonstrate that the interchange of vortices can be implemented by the rotation of the $U(1)$ phase $\Phi$.

%Numerically solving TBdG \eqref{eq:TBdG} is equivalent to doing the differential equation \eqref{eq:four}.
%Therefore, the eq.\eqref{eq:matrix} says that direct transition between pairs is forbidden by PHS, which can be checked by projecting the evolution state $\ket{\psi(t)}$ to instantaneous eigenstates.

To unveil the quasiparticle dynamics during the braiding of vortices, we compute the projection of the time-evolution of the encoded (initial) state $\ket{\varphi_{E_n}(\Phi(0))}$ onto the instantaneous eigenstates.
%\beq
%P^{(m)}_{n} (t) = \left| \braket{\varphi_{E_n}(\Phi(t))|\psi_{E_m}(t)} \right|^2.
%\eeq
Figures~\ref{fig:four_result}(a) and \ref{fig:four_result}(b) show the projections of $\ket{\psi (t)}$ onto the instantaneous eigenstates $\ket{\varphi_{E_n} (\Phi(t))}$, $| \braket{\varphi_{E_n}(\Phi(t))|\psi_{E_m}(t)}|^2$, where the initial states in (a) and (b) are $\ket{\varphi_{-E_1} (\Phi(0))}$ and $\ket{\varphi_{-E_2} (\Phi(0))}$, respectively. The quasiparticle during the braiding operation is composed of the instantaneous eigenstates with $-E_1$ and $+E_2$ ($-E_2$ and $+E_1$). The direct transition to the particle-hole symmetric eigenstate $+E_1$ ($+E_2$) is never observed, and the eigenstate with the energy $+E_1$ ($+E_2$) do not contribute to the dynamics of $-E_1$ ($-E_2$). This observation is understandable from the differential equation with the Berry connection matrix, 
\begin{equation}
i \partial_t \bm{C}(t) = \hat{B}(t) \bm{C}(t),
\label{eq:four}
\end{equation}
where  $\bm{C}(t)$ is $(C_{+E_1},C_{-E_1},C_{+E_2},C_{-E_2})$ and $\hat{B}(t)$ is
\begin{equation*}
\hat{B}(t) \equiv \left( \begin{array}{cc} \hat{0}_{2 \times 2} & \hat{A} _ {E_2 , E_1} \\ \hat{A} ^\dag _{E_2 , E_1} & \hat{0}_{2 \times 2} \end{array} \right).
\end{equation*}
As mentioned in Sec.~\ref{sec:selection}, Eq.~\eqref{eq:matrix} indicates that direct transition between pairs is suppressed by PHS, which can be confirmed by projecting the evolution state $\ket{\psi(t)}$ to instantaneous eigenstates.

As an important consequence, it can be seen from Figs.~\ref{fig:four_result}(a) and \ref{fig:four_result}(b) that the transition from $-E_1$ ($-E_2$) to $+E_2$ ($+E_1$) achieves $|\braket{\varphi_{+E_2}(\Phi(t))| \psi _{-E_1}(t)}|^2=|\braket{\varphi_{-E_1}(\Phi(t))| \psi _{-E_1}(t)}|^2=0.5$ [$|\braket{\varphi_{+E_1}(\Phi(t))| \psi _{-E_2}(t)}|^2=|\braket{\varphi_{-E_2}(\Phi(t))| \psi _{-E_2}(t)}|^2=0.5$] at $t\sim 3T/4$ before the interchange operation of vortices completes at $t=T$. Around $t\sim T/4$, the interchanging vortex moves across the branch cut associated with another interchanging vortex. It is also seen from Figs.~\ref{fig:four_result}(a) and \ref{fig:four_result}(b) that the transition probabilities, $|\braket{\varphi_{+E_2}(\Phi(t))| \psi _{-E_1}(t)}|^2$ and $|\braket{\varphi_{+E_1}(\Phi(t))| \psi _{-E_2}(t)}|^2$, tend to be saturated to $1$ before $t=2T$. These results indicate that the Hadamard and NOT gates may be accomplished  with small error even if the interchanged vortices do not precisely return to the initial positions.

%======================================
\begin{figure}
  \includegraphics[width=70mm]{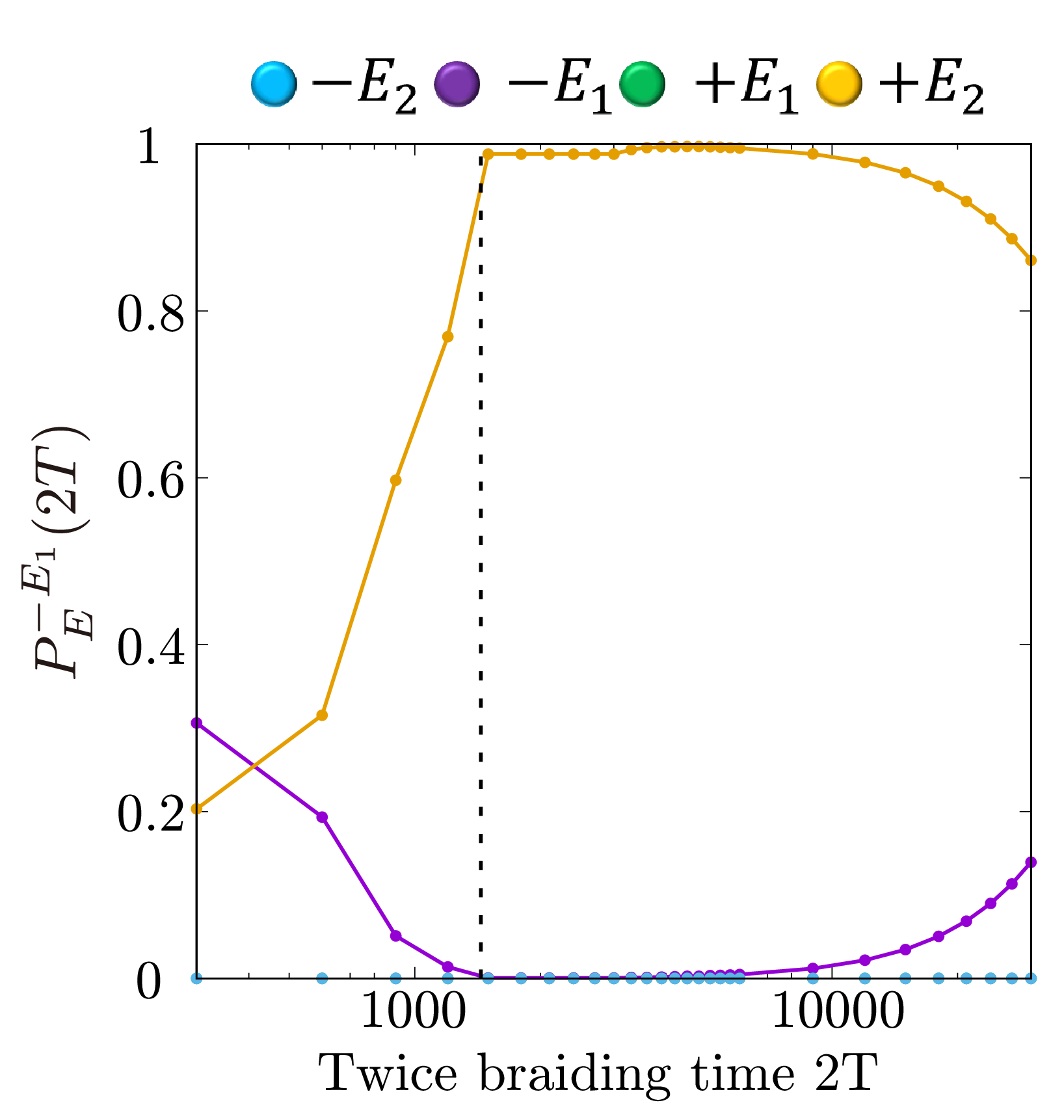}
\caption{Transition probabilities $P^{-E_1}_{E}$ as the function of the period of the twice braiding operation $2T$. The solid lines with different colors corresponds to the projection onto the eigenstates with $E=\pm E_2$ and $\pm E_1$, where the color is same as that in Fig.~\ref{fig:four_result}. The dashed line denotes $T=1440t^{-1}_0$ which is the braiding period taken in Figs.~\ref{four} and \ref{fig:four_result}.}
\label{fig:adiabatic_v1}
\end{figure}
%=====================================/

We also compute the transition probability of the encoded (initial) state $\ket{\psi_m(t=0)}=\ket{\varphi_{m}(\Phi(t))}$ to the initial state, 
\beq
P^{(m)}_n(t)\equiv | \braket{\varphi_{n}(\Phi(0))|\psi_{m}(t)}|^2.
\eeq
which is the projection of $\ket{\psi(t)}$ onto the encoded state. Figures~\ref{fig:four_result}(c) and \ref{fig:four_result}(d) show the projection of $\ket{\psi_m(t)}$ onto the quasiparticle states $\ket{\varphi_n(\Phi(0))}$ at $t=0$, $P^{(m)}_n(t)$. The interchange of vortices at $t=T$ leads to the equal superposition of the encoded $-E_1$ ($-E_2$) state with the $+E_2$ ($+E_1$) state. Another interchange operation completely transforms the encoded $-E_1$ ($-E_2$) state to the $+E_2$ ($+E_1$) state. The braiding dynamics of ``1'' and ``3'' vortices generates the transformation of Bogoliubov quasiparticles as $\hat{\eta}^{}_{+E_1}\rightarrow (\hat{\eta}_{+E_1} + e^{i \alpha} \hat{\eta}^{\dag}_{+E_2})/\sqrt{2}$ and $\hat{\eta}^{}_{+E_2}\rightarrow (\hat{\eta}_{+E_2} + e^{i\alpha} \hat{\eta}^{\dag}_{+E_1})/\sqrt{2}$. Hence, the twice interchange operation of the vortices transforms 
\beq
\hat{\eta}^{}_{+E_1} \hat{\eta}^{}_{+E_2} \rightarrow  \hat{\eta}^{\dag}_{+E_1} \hat{\eta}^\dag_{+E_2} ,
\eeq
which implies that braiding vortices generates the unitary evolution of the encoded state $\ket{00}\rightarrow\ket{11}$. 

%, $\ket{00}\rightarrow(\ket{00}+\ket{11})/\sqrt{2}$, and twice braiding is the completely transition to another degenerate groundstates, $\ket{00}\rightarrow\ket{11}$.
In Fig.~\ref{fig:adiabatic_v1}, we compute the transition probabilities $P^{-E_1}_{E}$ as the function of the twice braiding period $2T$.
Twice braiding of MZMs shows the transition from initial state to another degenerate states.
We find that the transition probabilities $P^{-E_1}_{+E_2} \approx 1, P^{-E_1}_{-E_1} \approx 0$,when the twice braiding period $2T$ satisfies $ 1000t^{-1}_0 \lesssim 2T \lesssim 10000t^{-1}_0$.
For $2T \lesssim 1000t^{-1}_0 $, $P^{-E_1}_{+E_2}$ remarkably decreases with increasing the speed of the braiding operation.
This is caused by the nonadiabatic transition to the higher vortex bound states.
For $2T > 10000t^{-1}_0$,  $P^{-E_1}_{+E_1}$ is gradually decreasing, while  $P^{-E_1}_{-E_1}$ is gradually increasing.
This implies that the braiding dynamics approaches the adiabatic limit, where the twice braiding period $2T$ is slower than the time sale of the energy splitting of MZMs.
Thus, in the adiabatic limit, $T \rightarrow \infty$, the braiding dynamics does not show non-Abelian statistics due to the effect of MZM hybridization.

The condition of the braiding operation period $T$ is $1.77~t^{-1}_0 \lesssim T \lesssim 2.7\times10^6~t^{-1}_0$. If one takes spin-orbit coupling constant as $\lambda=50~{\rm meV}$, the condition leads to $55.7{\rm fs}\lesssim T\lesssim 85.1{\rm ns}$. The lower bound represents the condition to avoid the nonadiabatic transition to the higher CdGM states, which is determined by the level spacing between CdGM states, $\delta E_{\rm CdGM}\sim \Delta^2_0/E_{\rm F}$. As an example, we take the values of the superconducting gap and the Fermi energy in Fe(Se,Te) as $\Delta_0=1.5~{\rm meV}$~\cite{Hanaguri474} and $E_{\rm F}\sim 10~{\rm meV}$~\cite{Rinotte1602372}, which leads to the level spacing $\Delta^2_0/E_{\rm F}\sim 0.23~{\rm meV}$. With these energy scales, the lower bound in Eq.~\eqref{eq:time} is estimated as the order of $\delta E^{-1}_{\rm CdGM}\sim 10{\rm ps}$. In contrast, the upper bound exponentially increases as a function of the ratio of the intervortex distance and the superconducting coherence length, and depends on an applied magnetic field. For the intervortex distance $100~{\rm nm}$ and the vortex core radius $10~{\rm nm}$ at $B=1~{\rm T}$ in Ref.~\cite{Machida2019}, the upper bound is estimated as the order of $0.48~{\rm s}$. These two bounds are well separated and different by several order of magnitude with realistic material parameters.

\section{Concluding remarks}

In conclusion, we have demonstrated non-Abelian statistics of vortices in two-dimensional class D topological superconductors. An $s$-wave superconductor with Rashba spin-orbit coupling, which we mainly consider here, is known as a prototype of class D topological superconductors hosting MZMs. However, such zero energy states are protected by a $\mathbb{Z}_2$ invariant and fragile against the hybridization due to the overlap of the wave functions.
In this work, we have numerically solved the TDBdG equation which incorporates the effect of the quasiparticle tunneling between neighboring vortices. This calculation does not assumes \textit{a priori} existence of MZMs.  In addition to $s$-wave superconductors with the Rashba spin-orbit coupling, we have also present the numerical simulation of non-Abelian braiding statistics in the heterostructure of an $s$-wave superconductor and a topological insulator in Appendix~\ref{sec:fu-kane}.

%We summarize the non-Abelian braiding statistics of MZMs and the symmetry consideration on the transition rules between splitting MZMs in Sec.~\ref{sec:2}.
In Sec.~\ref{sec:two}, we have demonstrated that the interchange of two vortices can be implemented by rotating the ${\rm U}(1)$ phase in trijunction systems and extracted the geometric phase from the numerical simulation of braiding dynamics. Owing to the suppression of direct transition between particle-hole symmetric eigenstates, the vortex-bound quasiparticles accumulate the nontrivial geometric phase $\pi/2$ in the interchange of two vortices. We have presented the upper and lower bound of the braiding time scale,
%, which is estimated as $\mathcal{O}(10^{-15})[s] \lesssim T \lesssim \mathcal{O}(10^{-9})[s]$.
within appropriate period the non-Abelian braiding statistics obeys $U(T)$ in Eq.~\eqref{eq:umatrix2} with $\vartheta = \pi/4$, irrespective of MZM hybridization.

%Such transition rule and geometric phase in the braiding dynamics of a two-vortex system are confirmed by numerically solving the TDBdG equation. 

%the braiding dynamics of vortices in $s$-wave superconductors with Rashba spin-orbit coupling using  We have confirmed the Braiding operator $U(T)$ in Sec.\ref{sec:two} and non-Abelian statistics in Sec.\ref{sec:four}:
%Geometric phase $\phi_{\rm geo}$ is close to $\pi/2$ in two-vortex system. 
%It say that the braiding dynamics satisfy $U(T)$ \eqref{eq:umatrix2} with $\vartheta = \pi/4$.
%Non-Abelian statistics is that after once braiding the initial state transit to the superposition between degenerate ground states and after twice it completely does another degenerate ground states.
%We have demonstrated the nature of non-Abelian statistics by solving TDBDG equation in Rashba model and Fu-Kane model.
%The order of braiding period is $\mathcal{O}(10^{-15})[s] < T < \mathcal{O}(10^{-9})[s]$, which is realistic time scale.

In Sec.~\ref{sec:four}, we have also performed the numerical simulation of braiding dynamics in topological superconductors with four vortices. It has been demonstrated that the quasiparticle dynamics hosted by vortex singularities obey the non-Abelian statistics. The numerical simulation shows that the twice interchange operation of two vortices gives rise the the transition of quasiparticle operators $\hat{\eta}^{}_{+E_1} \hat{\eta}^{}_{+E_2} \rightarrow \hat{\eta}^\dag_{+E_1} \hat{\eta}^\dag_{+E_2}$, which is the nature of non-Abelian statistics; $ \ket{00} \rightarrow \ket{11} $.From the numerical simulation, we have evaluated the adiabatic and non-adiabatic errors due to the energy splitting of MZMs and interactions to the higher energy CdGM states.
The braiding time scale is determined by the energy gap between MZMs and CdGM and the energy splitting of MZMs.
We find that the twice interchanging operation of the two vortices does not shows non-Abelian statistics in the region where the braiding period $T$ is close to the upper limit.
This leads to the serious error when the implementation of quantum gate using Majorana based qubits.

%We have demonstrated Majorana braiding dynamics using $s$-wave superconductors.
%This shows that the same result is expected in the $d$-wave superconductor.
We would like to point out some issues on the configuration of superconducting junctions for realizing braiding dynamics in experiments.
%First of all, we discuss $s$-wave superconductor case.
The ${\rm U}(1)$ phase $\Phi$ in the superconducting island (see Figs.~\ref{fig:two} and \ref{four_phase}) may be controlled by changing an external voltage or phase slip~\cite{astacoherent}.
In $s$-wave superconductors, however, Josephson current due to the phase difference between islands flows across domain walls, which may make such junction thermodynamically unstable. Here we would like to mention that Majorana fermions exist in a vortex core of a $d$-wave superconductor with an antisymmetric spin-orbit interaction and a nonzero magnetic field~\cite{satoPRL10}, which is protected by a topological invariant in spite of the presence of bulk gapless nodal quasiparticles. In high-$T_{\rm c}$ superconductors, it has been reported that integer and half-integer Josephson vortices are trapped in grain boundaries and tricrystal points~\cite{Kirtley1373,TsueiPRL73,KirtleyPRL76}. Although a $d$-wave superconductor offers a potential platform for realizing non-Abelian braiding statistics, the contribution of gapless nodal quasiparticles may significantly disturb the braiding dynamics of MZMs. 
%When MZMs is exchanged, MZMs are unstable due to the contribution of quasiparticle in nodal line.
In addition to superconducting heterostructures, the iron-based superconductor Fe(Se,Te) is a prime candidate of bulk topological superconductors hosting Majorana bound states in vortices~\cite{Machida2019,Wang333}. 
%We expect the candidate system is bulk system Fe(Se,Te).
%Fe(Te,Se) is the candidate for Majorana vortex in bulk system.
%The braiding method for bulk system is one of manipulating vortices used by spin-polarized STM~\cite{Wang333,SunPRL116}.
Motion of vortices hosting MZMs might be manipulated by using a spin-polarized STM tip~\cite{SunPRL116} or magnetic force microscopy \cite{benjamin}.
%We expect it will give similar results.
Another important issue, which has not been discussed here, is the decoherence of the Majorana qubit due to interaction with thermal environment~\cite{HenrikSPP,ChengPRB85}, where the fermion occupation in the Majorana qubit may leak into the thermal bath. How nodal quasiparticles and thermal excitations disturb non-Abelian braiding dynamics remains issues for future research.

Lastly, we would like to mention that the numerical method presented here is generalizable to the other systems, including superconducting nanowires, planar Josephson junctions, and the other topological classes with symmetry-protected MZMs~\cite{satoE14,liuPRX14,haim}. In particular, the topological superconducting phase in proximitized semiconductor nanowires provides more promising platform for topological quantum computation. However, the braiding dynamics in the T junction network of nanowires may suffer intrinsic and extrinsic disturbances. This includes a non-uniformity of the proximitized superconducting gap introducing geometric dependence to the dynamical phase, thermally excited quasiparticles, the fluctuation of fermion parity to the ancilla nanowires, and gate voltage fluctuations. 
%Then fundamental questions arise. How can the coherence of Majorana qubits and quantum gates be protected against intrinsic and extrinsic disturbances? Braiding dynamics in the T junction network of nanowires may suffer 
%a non-uniformity of the proximitized superconducting gap may introduce geometric dependence to the dynamical phase, and the braiding MZMs in the T junction network of nanowires may suffer the fluctuation of fermion parity to the ancilla nanowires. 
If a system maintains an anti-unitary symmetry $\mathcal{T}^2=-1$, processing quantum information with Majorana Kramers pairs is sensitive to local perturbations that cause local mixing of degenerate ground states via time-dependent symmetry-preserving coupling to bulk quasiparticles~\cite{wolms,gao}. All these may be harmful for the coherence of Majorana qubits and quantum gates. The {\it ab initio} simulation on such decoherence remains as future works. In addition, another fundamental question arises. How does the particle number conservation affect the braiding dynamics and the performance of topological quantum computation with Majorana zero modes? In accordance with the number conserving theory~\cite{lin2017effect,lin2018towards}, the dynamics of Bogoliubov quasiparticles is inevitably accompanied by the dynamics of Cooper pairs. 
%Recently, the number conserving theory including the results of BCS mean-field is featured ~\cite{LapaPRL124,KnappPRB101}. We believe that the results of this paper should be demonstrated in particle number conserving theory and will attempt to construct the number conserving model like BdG equation. ... 
This is also a fundamental and important issue to be addressed in the future.

\begin{acknowledgments}
We thank Y. Nagai for valuable technical advice on numerical calculation and Y. Maeno and K. Takase for fruitful comments on the experimental setup of trijunction network. This work was supported by a Grant-in-Aid for Scientific Research on Innovative Areas ``Topological Materials Science'' (Grants No.~JP15H05852, No.~JP15H05855, and No.~JP15K21717), and ``J-Physics'' (JP18H04318), and ``Quantum Liquid Crystals (JP20H05163)'' from JSPS of Japan, and JST CREST Grant Number JPMJCR19T5, Japan, and JSPS KAKENHI (Grant No.~JP16K05448, No.~JP17K05517, and No.~JP20K03860). T.S. was supported by a JSPS Fellowship for Young Scientists.
\end{acknowledgments} 

%20H05163 QLC (T.M.)
%20K03860 Kiban-C (T.M.)

%%%%%%%%%%%%%%%%%%%%%
%Using successive substitution method, we get the \( \bm{C}(t) \) :
%\begin{equation*}
%\begin{split}
%\bm{C}(t) &= \bm{C}(0) +(-i) \int_0^t dt' \hat{B}(t')\bm{C}(t')
%\\
%&=\left(\hat{1} + (-i) \int_0^t dt' \hat{B}(t')\right) \bm{C}(0)s
%\\
%& \qquad+ (-i)^2 \int_0^{t} dt' \hat{B}(t') \int_0^{t'} dt'' \hat{B}(t'') %\bm{C}(t'')
%\\
%&=...
%\\
%&=\hat{T}  \exp \left[ -i \int_0^t dt' \hat{B}(t') ) \bm{C}(0) \right] , 
%\end{split}
%\end{equation*}
%where \( \hat{T} \) is time order operator.

\appendix

\section{Non-Abelian statistics in the Fu-Kane model}
\label{sec:fu-kane}
In this appendix, we demonstrate non-Abelian statistics in the Fu-Kane model~\cite{fuPRL100}. 
The model comprises a topological insulator with an $s$-wave superconductor layer, yielding a proximitized two-dimensional Dirac fermion. This  can be a prototypical system of a topological superconductor hosting MZMs.
%This numerical simulation cannot compare Rashba model with the braiding stability, but it is worthwhile to simulate braiding dynamics with different models.
%The surface of topological insulator is the twodimensional Dirac fermion $+$ s-wave superconductor because cooper pairs tunnel into that by proximity effect.
%The low energy spectrum is spinless $p$-wave superconductor hosting MZMs.
%The similar setup occurs at the interface of Fu-Kane model.

\begin{figure}[t!]
    \centering
    \subfloat{\includegraphics[width=8cm,height=4cm]{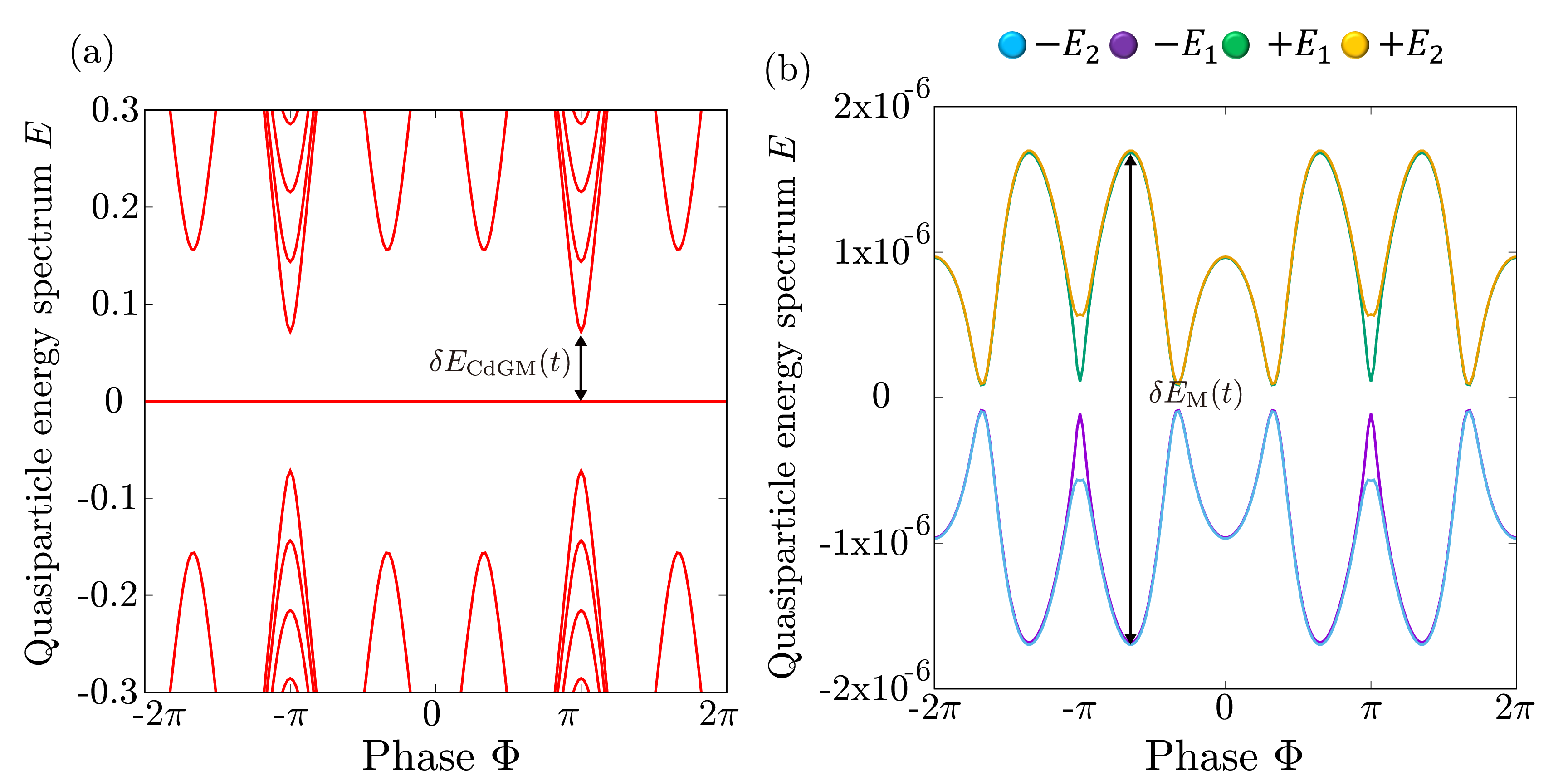}}
    \\
    \subfloat{\includegraphics[width=7cm,height=5cm]{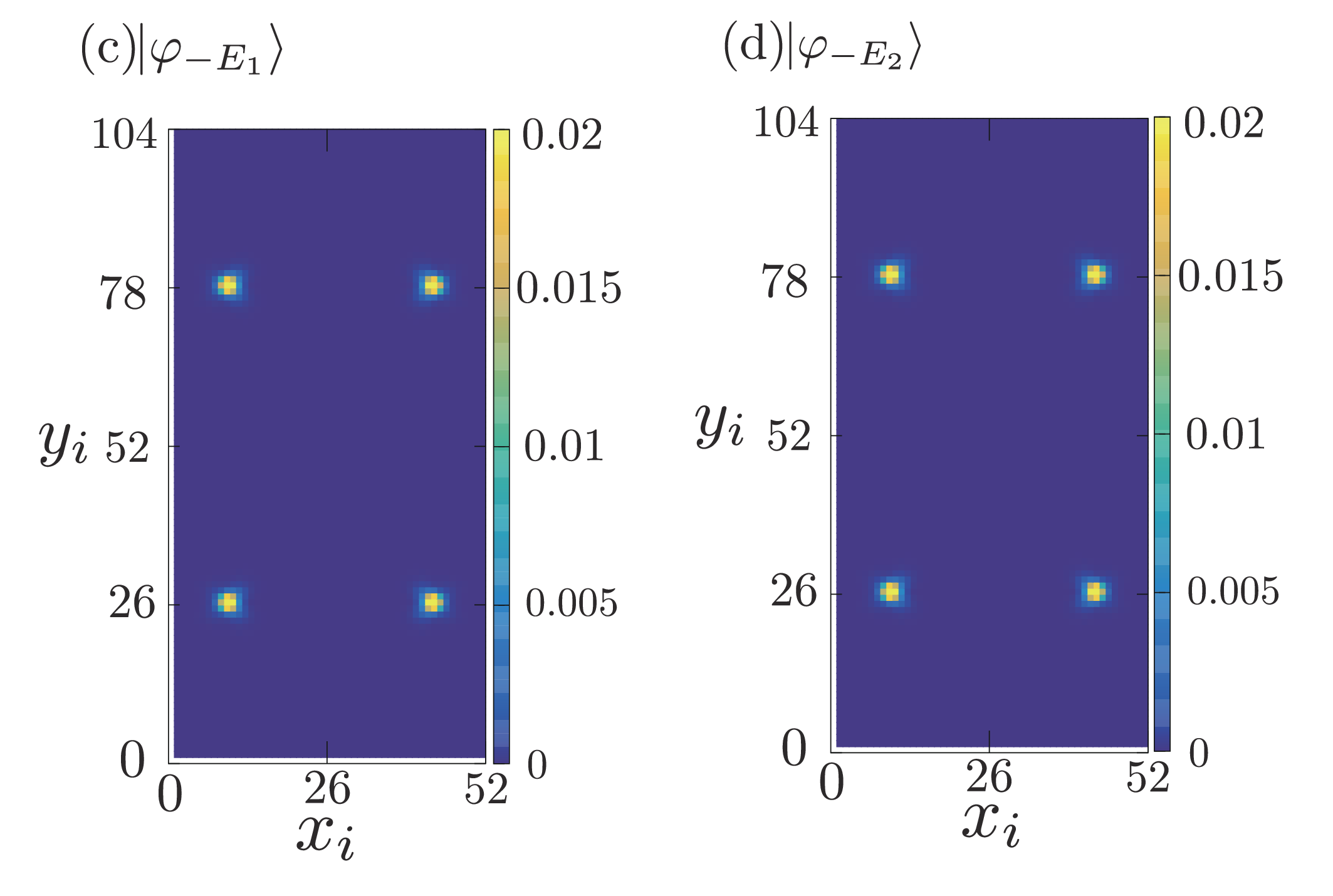}}
    \caption{(a) The lower bound of the braiding period $T$ is determined by the inverse of the minimum splitting between MZMs and the higher CdGM states around $\Phi=\pm\pi$, $\delta E^{-1}_{\rm CdGM}(t)$. (b) Instantaneous eigenenergies of Majorana band, where the inverse of the maximum splitting occurs around $\Phi=\pm 16\pi/25$, $\delta E^{-1}_{\rm M}(t)$, determines the upper bound of $T$. (c,d) Spatial profiles of the quasiparticle wave function with the energy $-E_1$ and $-E_2$,  $\sum _{\sigma}[|u_{{\bm i},\sigma}|^2+|v_{{\bm i},\sigma}|^2]$, where the $U(1)$ phase is set to be $\Phi=-2\pi$.}
    \label{fig:fu_setup}
\end{figure}

Here we start with the effective Hamiltonian of the Fu-Kane model, $\mathcal{H}_{\rm FK}$~\cite{marchandPRB86,liuPRB92}, which is written in the momentum space as
\begin{equation}
\mathcal{H}_{\rm FK} =
\begin{pmatrix}
g_{\bm{k}} - \mu& M_{\bm{k}} & \Delta_{\bm{k}} & 0\\
M_{\bm{k}} & -g_{\bm{k}} - m \sigma_z & 0 & 0 \\
\Delta^*_{\bm{k}} & 0 & -g^{\rm tr}_{-\bm{k}} + \mu & -M_{\bm{k}}\\
0&0&-M_{\bm{k}} & g^{\rm tr}_{-\bm{k}} + m\sigma_z
\end{pmatrix}.
\label{eq:HFK}
\end{equation}
where 
\begin{gather}
g_{\bm k} = 2\lambda (\sigma_y\sin k_x-\sigma_x\sin k_y), \\
M_{\bm k} = 2\tau (2-\cos k_x - \cos k_y),\\
\delta_{\bm{k}} = i \sigma_y \Delta_0.
\end{gather}
The Hamiltonian acts on an eight-component Nambu spinor 
$(c^{\dag}_{\uparrow,1,{\bm k}},c^{\dag}_{\downarrow,1,{\bm k}},c^{\dag}_{\uparrow,2,{\bm k}},c^{\dag}_{\downarrow,2,{\bm k}},
c_{\downarrow,1,-{\bm k}},-c_{\uparrow,1,-{\bm k}},c_{\downarrow,2,-{\bm k}},-c_{\uparrow,2,-{\bm k}})$, where $c^{\dag}_{\sigma,i,{\bm k}}$ is a creation operator of a fermion with spin $\sigma = \uparrow,\downarrow$ and momentum ${\bm k}$ on the surface $i=1,2$ of the topological insulator. The diagonal blocks describe the gapless surface Dirac fermions on the two surfaces of the topological insulator, and $M_{\bm{k}}$ generates an energy gap in all the Dirac nodes except those at $\bm{k} = (0,0)$.
An $s$-wave superconductivity is induced in one of the surfaces, the surface ``1'', by proximity effect. To study vortex dynamics, we transform the effective Hamiltonian in the momentum space to a two-dimensional square lattice in the real space. The Hamiltonian in Eq.~\eqref{eq:HFK} includes an exchange interaction on the surface ``2'', $m\sigma_z$, which induces a gap in the surface states. The term is necessary to remove unwanted gapless excitations from the surface ``2''~\cite{liuPRB92}.

At $\mu=0$ the Fu-Kane model maintains the chiral symmetry with $\gamma^5$, and the MZMs have a well-defined chirality. The symmetry prohibits the hybridization of MZMs with same chirality, and rigidly protects MZMs with $\mathbb{Z}$ topological invariant rather than $\mathbb{Z}_2$~\cite{chengPRB82}. The index theorem~\cite{WeinbergPRD24,FukuiJPSJ} also ensures that $N$ singly-quantized vortices host $N$ MZMs, and the chiral symmetry prevents MZMs from hybridization. The deviation from $\mu=0$ breaks the chiral symmetry, and gives rise to the hybridization. The MZMs bound at vortices start to form a band structure.

\begin{figure}[t!]
    \centering
	\includegraphics[width=8cm,height=4cm]{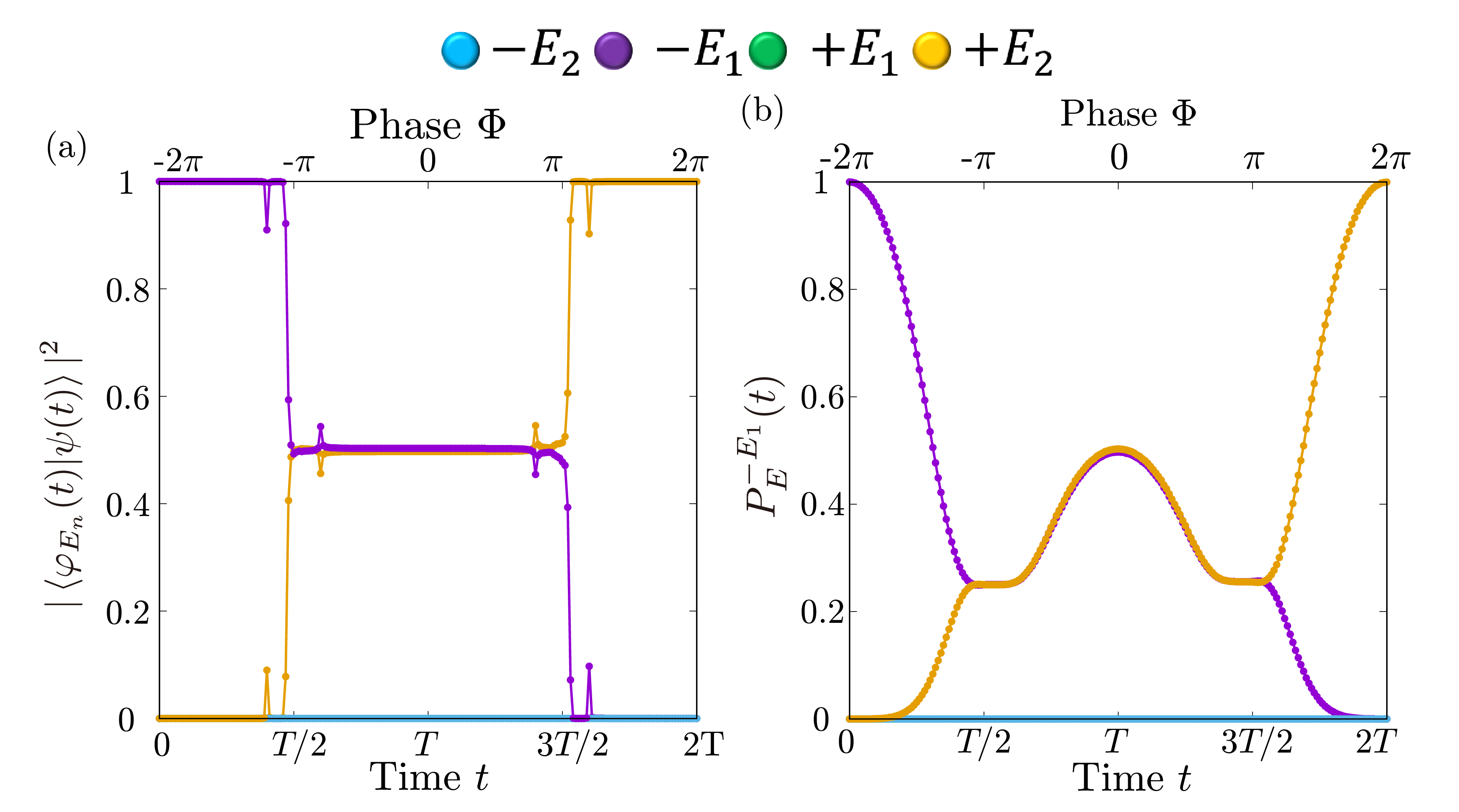}
    \caption{Projection of $\ket{\psi (t)}$ onto the instantaneous eigenstates (a) and the encoded state $\ket{\varphi_{-E_1} (\Phi (0))}$ (b). where the period $T$ is fixed to $T=7500\tau^{-1}$. The twice interchange operation of two vortices generates the transition of the quasiparticle states with $-E_1$ to the eigenstate with $+E_2$.}
    \label{fig:fu_results}
\end{figure}

Here we discuss braiding dynamics in the Fu-Kane model with four vortex singularities, where  the configuration of four vortices is same as superconducting junctions in Fig.~\ref{four_phase}. For the numerical simulation, we set the parameter $\tau=0.9$, $\mu=4.8$, $\lambda=2.0$, $m=0.5$, and $\Delta_0=1.6$. We consider a system with four vortices, which hosts two particle-hole symmetric vortex bound states, $\pm E_1$ and $\pm E_2$, as a consequence of the hybridization of four MZMs. Figures~\ref{fig:fu_setup}(c) and (d) show the wave function of the lowest $( E_1 / \tau = 9.616 \times 10^{-7} )$ and second lowest $(E_2 / \tau = 9.678\times 10^{-7})$, $\sum _{\sigma}[|u_{{\bm i},\sigma}(t)|^2+|v_{{\bm i},\sigma}(t)|^2]$.

For numerical simulation of braiding vortices, we start with the initial state, $\ket{\varphi_{-E_1}}$. The braiding period is determined in the same manner as that in the main text, i.e., the appropriate period must obey $1.40 \times 10 \tau^{-1}\lesssim T \lesssim 2.94 \times 10^5\tau^{-1}$. The lower bound is set to prevent MZMs from non-adiabatic coupling to other quasiparticle states with higher energies, while the upper bound is necessarily to protect non-Abelian braiding dynamics from errors due to dynamical phase accumulation. Numerical simulation demonstrates non-Abelian braiding dynamics with high accuracy as long as the braiding operation satisfies the condition. 
%We set the initial state as $\ket{\varphi_{-E_1}}$ and the braiding period $T$ is $T=7500 \tau^{-1}$:$1.40 \times 10 < T/\tau < 2.94 \times 10^5$.
%
%The numerical result is expected that $\ket{\varphi_{-E_1}}$ transits to $\ket{\varphi_{+E_2}}$ like main section.

\begin{figure}[t!]
    \centering
    \subfloat{\includegraphics[width=70mm]{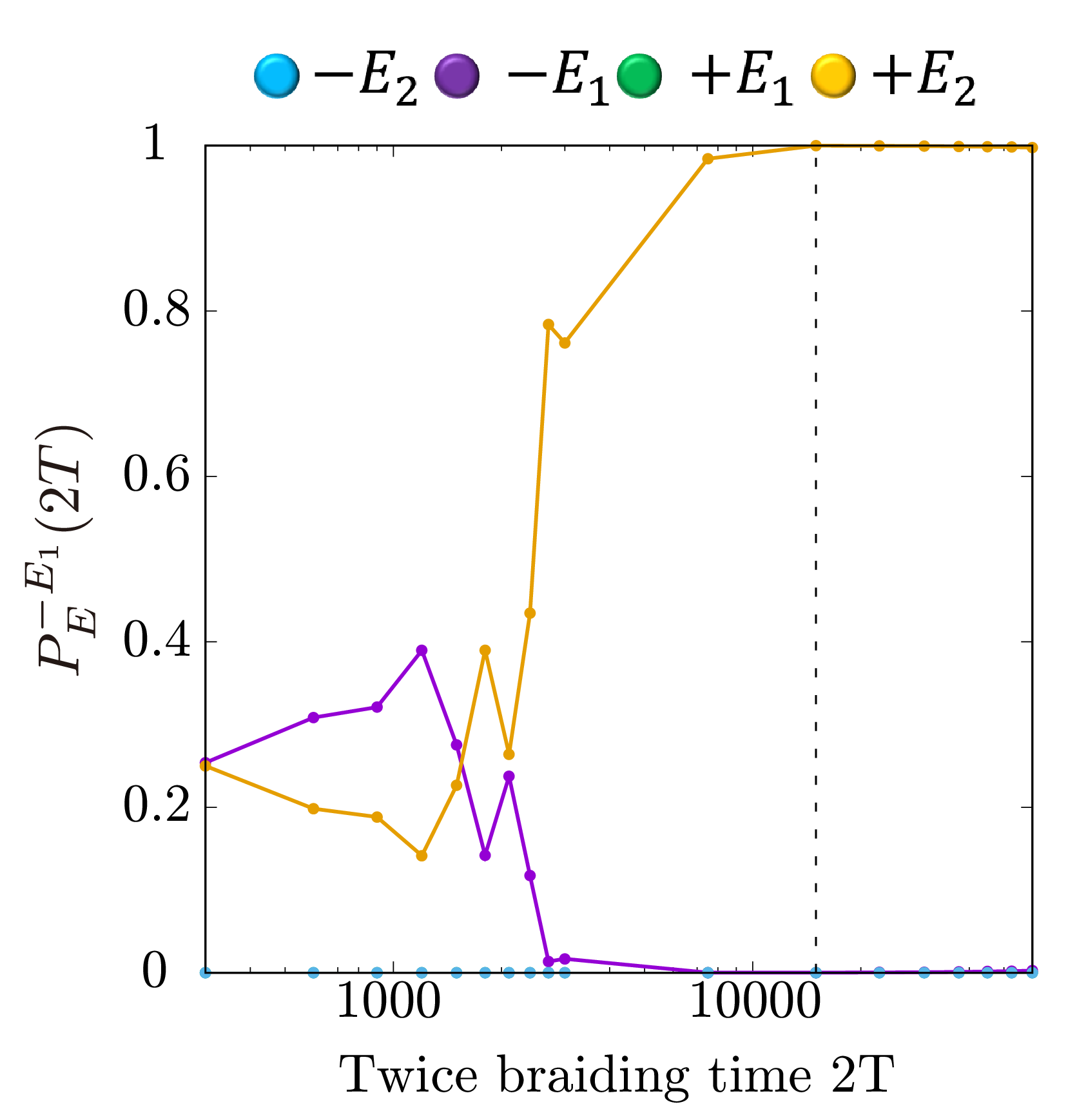}}
    \caption{Transition probabilities $P^{-E_1}_{E}$ as the function of the period of the twice braiding operation $2T$. The numerical simulation with $T=7500t^{-1}_0$, which is shown in Fig.~\ref{fig:fu_results}, is denoted by dashed line.}
    \label{fig:fu_T}
\end{figure}

In Fig.~\ref{fig:fu_results}(a), we plot the projection of the time-evolved eigenvector $\ket{\psi (t)}$ onto the instantaneous eigenstates and the encoded state $\ket{\varphi_{-E_1} (\Phi (0))}$ [Fig.~\ref{fig:fu_results}(b)]. Figure~\ref{fig:fu_results}(b) also shows the projection of  $\ket{\psi (t)}$ onto  the encoded state $\ket{\varphi_{-E_1} (\Phi (0))}$, revealing the transition from initial state to another degenerate ground state: $P^{-E_1}_{E} (T)$. Similarly with the numerical simulation in the Rashba model, both figures demonstrate that the interchange of two vortices in four vortices gives rise to the transition from $-E_1$ to $+E_2$. The time-evolved eigenvector is composed of equal contributions of the $-E_1$ and $+E_2$ instantaneous eigenstates around $t=T/2$. This implies that one of the interchanging vortices moves across the branch cut around $t=T/2$ and experiences the abrupt $2\pi$-phase rotation. 
%
%There is the different point to compare to main section.
%We see that around $t \sim T/2$  the interchanging vortex move across the brunch cut, while it does the brunch cut around $t \sim T/4$ in main section.
%If the distance between the vortices is small and strongly hybridized, it is not necessarily to link a state transition and across the branch cut.
%Figure~\ref{fig:fu_results}(b) shows transition from initial state to another degenerate ground state:$P^{-E_1}_{E} (T)$. $P^{-E_1}_{\pm E_1}$ is purple and green line and $P^{-E_1}_{\pm E_2}$ is yellow and blue lines. This result is consistent with main result: $\ket{\varphi_{-E_1}}$ transits to $\ket{\varphi_{+E_2}}$ by interchanging two vortices.
%

In Fig.~\ref{fig:fu_T}, we compute the transition probabilities $P^{-E_1}_E$ as the function of the twice braiding period $2T$. Similarly with Fig.~\ref{fig:adiabatic_v1}, $P^{-E1}_{+E_2}$ remarkably decreases for $2T \lesssim 10000 \tau^{-1}$. This deviation is attributed to the nonadiabatic transition from MZMs to the higher energy quasiparticle states, i.e., nonadiabatic interaction to environment induces the decoherence of the Majorana-based qubit. 
$P^{-E_1}_{-E_1}$ is not gradually decreasing in Fig.~\ref{fig:fu_T}.
Our numerical results in the Fu-Kane model does not show the deviation of $P^{-E1}_{+E_2}$ in the adiabatic limit, $T\gg \delta E^{-1}_{\rm M} \sim 10^{5}$. The numerical simulation with the much slower braiding operation, $T \gg 10^{5}$, is required to realize the adiabatic errors of non-Abelian braiding statistics in the Fu-Kane model.

Let us discuss the time scale of the braiding dynamics in iron-based superconductors Fe(Se,Te). The superconducting gap $\Delta$ of Fe(Se, Te) is observed as $\Delta \approx 1.5~{\rm meV}$ ~\cite{Hanaguri474} and its Fermi energy $E_{\rm F}$ is $E_{\rm F} \approx 10~{\rm meV}$~\cite{Rinotte1602372}. By using these values, the typical energy spacing between the CdGM states is estimated as $\delta E_{\rm CdGM}\approx \Delta^2/E_{\rm F}= 0.23~{\rm meV}$. The level spacing determines the lower bound of the braiding period $T$.
In contrast, the upper bound exponentially increases as a function of the ratio of the intervortex distance and the superconducting coherence length, and depends on an applied magnetic field. For the intervortex distance $100~{\rm nm}$ and the vortex core radius $10~{\rm nm}$ at $B=1~{\rm T}$ in Ref.~\cite{Machida2019}. MZM hybridization can be approximated as $\delta E_M \sim \Delta \exp ( -R/ \xi)/ R^{1/2}$ since topological surface states of Fe(Se, Te) is observed around $\Gamma$ point~\cite{Zhang182}. As shown in Fig.~\ref{fig:fu_setup}(a), the minimum gap from MZMs in instantaneous energy spectrum is $\min [ \delta E_{\rm CdGM}(t) ] = 6.56\times10^{-2}~{\rm meV}$. The condition of the braiding period $T$ is given by $10~[{\rm ps}] \lesssim T \lesssim 0.48 ~[{\rm s}]$. The braiding time scale might be feasible.

%We expect that the adiabatic limit is estimated by increasing the braiding period $T$.

%%%%%%%%%%%%%%%%%%%%%%%%%%%%%%%%%%%%%%%%%%%%%%%%%%%%%%%%%%%%%%%%%%%%%%%%%

\bibliographystyle{apsrev4-1_PRX_style.bst} % Use this style to include Titles in the Bibliography
\bibliography{majorana.bib}
\end{document}